\documentclass{article}

\usepackage{arxiv}

\usepackage[utf8]{inputenc} 
\usepackage[T1]{fontenc}    
\usepackage{hyperref}       
\usepackage{url}            
\usepackage{booktabs}       
\usepackage{amsfonts}       
\usepackage{nicefrac}       
\usepackage{microtype}      
\usepackage{lipsum}
\usepackage{graphicx}
\usepackage{natbib}
\usepackage{multirow}
\graphicspath{ {./images/} }

\title{High Frequency Radar Observing System Simulation Experiment in the Western Mediterranean Sea: a Lagrangian assessment approach.}

\author{Jaime Hernández-Lasheras \\
        Balearic Islands Coastal Observing and Forecasting System (SOCIB), Palma, Spain \\    CMCC Foundation - Euro-Mediterranean Center on Climate Change, Italy.\\
        jhlasheras@gmail.com
 \And
        Alejandro Orfila \\
        IMEDEA (CSIC-UIB) - Instituto Mediterraneo de Estudios Avanzados. Esporles, Spain
 \And
        Alex Santana \\ 
        Balearic Islands Coastal Observing and Forecasting System (SOCIB), Palma, Spain
 \And
        Ismael Hernandez-Carrasco \\
        IMEDEA (CSIC-UIB) - Instituto Mediterraneo de Estudios Avanzados. Esporles, Spain 
 \And 
        Baptiste Mourre \\ 
        Balearic Islands Coastal Observing and Forecasting System (SOCIB), Palma, Spain \\  IMEDEA (CSIC-UIB) - Instituto Mediterraneo de Estudios Avanzados. Esporles, Spain
        }

\begin{document}
\maketitle
\begin{abstract}

The impact of the expansion of a high-frequency radar (HFR) system in a dynamic coastal area (the Ibiza Channel in the Western Mediterranean Sea) is evaluated through an Observing System Simulation Experiment (OSSE). The installation of two new antennas in the Iberian Peninsula would complement the existing ones in the islands of Ibiza and Formentera, providing surface currents observations of the full channel. Two different configurations of the same model, validated to give realistic simulations, are used: i) a Nature Run (NR) which is considered as the real ocean state and that is used to generate pseudo-observations, and ii) a Control Run (CR) in which the pseudo-observations are assimilated. The OSSE is first validated by comparison against a previous Observing System Experiment (OSE). The impact of the new antennas for forecasting surface currents is evaluated in two different periods with different levels of agreement between NR and CR. The HFR expansion is found to contribute to significantly correct the circulation patterns in the Channel, leading to surface merdional velocity error reductions up to 19\%. The effects on the transport in the area are also analyzed from a Lagrangian perspective, showing that DA can help to better represent the Lagrangian Coherent Structures present in the NR and constrain the ocean dynamics.

\end{abstract}

\keywords{Data Assimilation \and High-frequency radar \and Operational oceanography \and Regional modelling \and Lagrangian validation}

\section{Introduction}
\label{sect:intro}

Observations, models and data assimilation (DA) are the three key elements in operational oceanography. Combining them in an optimal way and bridging synergies between the different research communities is key to advance our knowledge of the oceans and be able to answer to societal needs for a sustainable development \citep{Ryabinin2019TheDevelopment, Visbeck2018OceanFuture}.

 In this sense, Ocean Observing Systems (OOS) play a key role, and numerous efforts have been made all over the world to enhance its development and strengthen the collaboration within the scientific community \citep{Sloyan2019EvolvingCoordinatio,Moltmann2019ATechnologies,deYoung2019An2030}. In particular, in Europe, several initiatives have been made or are ongoing to provide better answers to science and to societal challenges (e.g., CMEMS programme, Jerico and Eurosea projects) \citep{Farcy2019TowardInfrastructure,LeTraon2019FromPerspective,Tintore2019ChallengesSea}.

The rising capabilities of remote sensing and the development during the last decades of \textit{in-situ} observing programs such as Argo \citep{LeTraon2013}, allowed a better understanding of ocean processes at multiple scales. In coastal areas, Regional Ocean Observing Systems (ROOS) are nowadays providing near real time observations combining observations from moored instruments, periodic cruises,  autonomous vehicles, Lagrangian platforms or high-frequency radars (HFR), among others.  

Numerical models provide a complete view of the three dimensional structure of the ocean. However, they are inevitably affected by errors from parametrization of non resolved physical processes, discretization issues, or the lack of accurate forcing. To improve reliability, numerical models for operational purposes should be fed with observations through data assimilation.  Observing System Experiments (OSEs) assimilating data are numerical experiments that can be designed to evaluate the capability of specific observing systems and eventually correct model forecasts simulations. Similarly, the  potential impact of observing systems has to be evaluated to help design these systems. Observing System Simulation Experiments (OSSEs) can be performed to help to optimally design an OOS or a future campaign \citep{Kourafalou2015}.

OSSEs were first developed for the atmospheric science community, and over time, specific design criteria have been developed to ensure the realism of the assessments performed, as defined in \cite{atlas1997atmospheric}.  In ocean studies, multiple OSSEs had been done, however, most of them did not use a full-fledged DA system approach for the evaluation. Generally, Kalman filters, empirical orthogonal functions (EOF) based or different interpolation methods were used to map the observations and reconstruct the ocean state \citep{Guinehut2004CombiningObservations, ballabrera2007observing, sakov2008osse}. Following the procedure established for atmospheric studies \citep{hoffman2016future}, \cite{Halliwell2014RigorousMexico} applied them for the first time in the ocean, and in the last years, several studies have been performed following the criteria exposed in that work, as we will do here. For instance,  \cite{Gasparin2019RequirementsExperiments} performed an evaluation of the influence of the future deep Argo float network, and \cite{Benkiran2021AssessingMethods} assessed the impact of the assimilation of data from the future SWOT satellite mission in a global-high-resolution model. \cite{aydougdu2016assimilation, aydougdu2018osse} also used this same approach to perform a design of observing systems in the Marmara and the Adriatic seas respectively.

In the fraternal twin OSSE approach employed, two models are required: (i) one, hereinafter referred to as Nature Run (NR),  which is considered to represent the true ocean and that will be used for validation and to extract the pseudo-observations, and (ii) the model we would like to correct with the assimilation of such pseudo-observations. To be credible, the OSSE should satisfy the following design criteria and rigorous evaluation steps: (a) The models should be validated to give realistic simulations and the pseudo-observations generated in a way that resemble the real ones, including the observation errors, that need to be specifically added. (b) The validation should be performed by comparison to a previous OSE where real observations are assimilated. The same observations should be assimilated in both experiments, except for the fact that, in the OSSE, the pseudo-observations are synthetically generated from the NR.  (c) If the impact assessment is consistently the same, we consider the OSSE to be validated.

In the OSSE framework the ocean state is fully known. This permits to assess the impact in regions that normally are not sampled or to experiment additional validation techniques. For instance, we can use Lagrangian techniques for the assessment of transport and the ocean dynamics, such as the Lagrangian Coherent Structures (LCS) obtained from the Finite Size Lyapunov Exponents (FSLE) \citep{dOvidio2004MixingExponents,Hernandez-Carrasco2011HowDynamics}. Ridges of FSLE field reveal LCS, which act as transport barriers. These LCS have been proven to be useful to understand ecological processes, such as nutrients distribution, or oil-spill and search and rescue operations \citep{Hernandez-Carrasco2018EffectSeas,lekien2005pollution, shadden2005definition}.

The validation of these Lagrangian techniques is generally limited. LCS computed from model simulations can be compared to those calculated from geostrophic currents derived from altimetry products \citep{https://doi.org/10.1029/2017JC013613} which suffer limitations when approaching the coast \citep{Vignudelli2019SatelliteZone,Pascual2013}. Also from HFR measured currents \citep{Hernandez-Carrasco2018EffectSeas}, that are limited to cover small coastal areas. The validation of LCS can be performed by comparison with active tracers, as chlorophyll filaments \citep{Lehahn2007StirringData,Hernandez-Carrasco2018EffectSeas}, which by their nature can not depict the full structures present in the ocean; SST fronts \citep{dOvidio2004MixingExponents}, which are inferred from satellite products that can be affected by clouds; or fish stock concentrations \citep{diaz2022singularities}, tracked seabirds or marine predators \citep{kai2009top}, that are difficult to monitor. Here we will take profit of the full ocean state knowledge supposed in the OSSE approach. The use of a NR model for the validation of the LCS computed from the model simulations implies a step forward to address the question of how data assimilation can help models to correct the circulation, especially in coastal areas.

This study complements the work presented in \cite{Hernandez-Lasheras2021EvaluatingModelling}, where a series of OSEs were performed to evaluate the impact of HFR DA on the correction of surface currents in the Ibiza Channel (IC).  In that work, it was shown that using HFR observations together with satellite observations (altimetry and sea surface temperature) and Argo temperature and salinity profiles increased the model's capability to forecast surface currents. In this work we will use the same set-up, from the Western Mediterranean OPerational modelling system (WMOP). 


The paper is structured as follows; Section 2, presents the data, the general set-up of the OSSE and the Lagrangian approach that will be followed. In Section 3, the OSSE performance is presented as well as the performance in the Eulerian and Lagrangian frameworks. The next sections discuss the main results and conclusions of the work.


\section{Data and Methods}

\subsection{Study Area and HFR system}

Our region of study is the Ibiza Channel (IC), in the Western Mediterranean Sea. The modelling area spans from Gibraltar strait in the west to Sardinia and Corsica in the east. The IC is a passage of water between the Iberian peninsula and the island of Ibiza, in the western Mediterranean Sea \citep{Pinot1994,Pinot1995}. It is a choke point between the saltier waters from the north, that generally flow along the coast, and the incoming fresher waters from the south \citep{Heslop2012}. 

\begin{figure}[!ht]
  \centering
  \includegraphics[width=.98\linewidth]{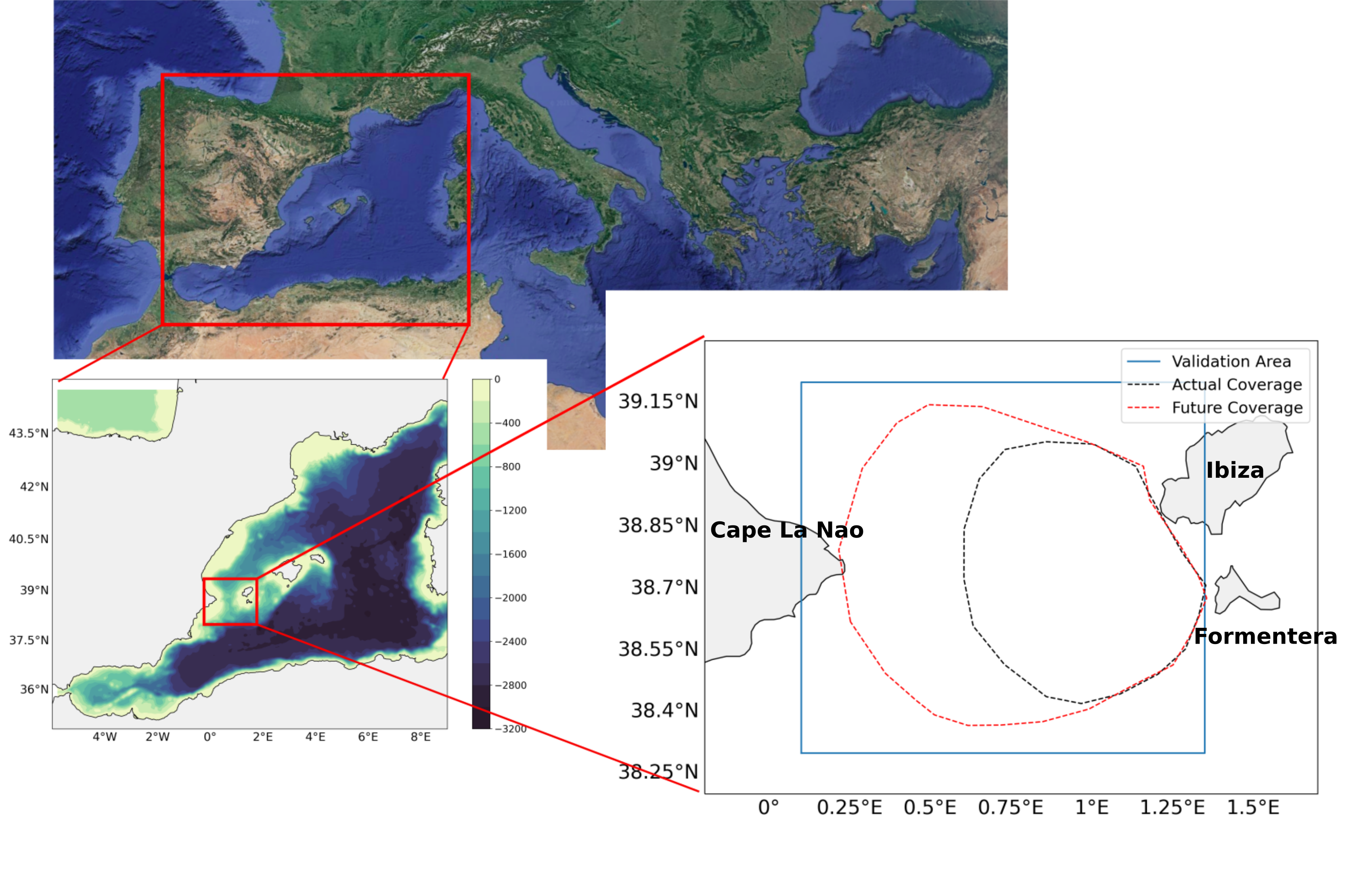}
    \caption{ WMOP model domain and bathymetry in the Western Mediterranean Sea. HFR present (black dashed) and potentially future (red dashed) coverage area  over which observations have been simulated. The blue rectangle represents the area selected for the validation.}
  \label{fig:coverage_area}
\end{figure}

As part of the Mediterranean HFR observing network (\citep{Lorente2022,Reyes2022}, SOCIB operates since 2012 two CODAR HFR antennas in the islands of Ibiza and Formentera, measuring surface currents in an area up to 80 km far off the coast \citep{tintore2012hf}. Here we  evaluate the potential impact that a couple of new antennas potentially installed in the Iberian peninsula, in the eastern part of Cape la Nao (Fig. \ref{fig:coverage_area}), would have on the WMOP modelling system. By installing these new antennas, the IC HFR system will expand its coverage area covering the entire channel. The area, shown in Fig. \ref{fig:coverage_area} (red dashed line), is considered the most likely coverage in terms of total velocities (u-v) that a couple of antennas installed in the western side of the Channel would provide together with the present system.

\subsection{OSSE set-up: Simulations}
\label{met:nr}

The OSSE perspective requires two model simulations. Here we used a fraternal twin OSSE system approach \citep{Halliwell2014RigorousMexico}, in which two different configurations of the same model are employed: i) a Nature Run (NR),  which is considered to represent the true ocean and that will be used for validation and to extract the pseudo-observations, and (ii) a Control Run (CR) model in which we will evaluate the impact of assimilating different observing sources. We use the WMOP modelling system \citep{Juza2016,Mourre2018AssessmentSea,Hernandez-Lasheras2018DenseSardinia}, which is a regional configuration of ROMS \citep{Shchepetkin2005} for the Western Mediterranean Sea coupled with a Local Multimodel Ensemble Optimal Interpolation (EnOI) data assimilation engine. It spans from Gibraltar Strait in the west to Corsica and Sardinia straits in the east, with a horizontal resolution around 2 km and 32 vertical terrain-following sigma levels (resulting in a vertical resolution between 1 and 2m at the surface). The NR and CR models share the same configuration, parametrization and atmospheric forcing, but differ in their initial state and boundary conditions. The CR is a free-run hindcast simulation developed and evaluated in \cite{Mourre2018AssessmentSea} and  \cite{Aguiar2019Multi-platformActivity}.  It uses the Copernicus Marine Forecasting System for the Mediterranean Sea (CMEMS MED-MFC), with a 1/16$^\circ$ horizontal resolution \citep{CMEMS}, as initial and lateral boundary conditions. This initial and boudary condition configuration is the same as that  used in the companion OSE that we will use for validation \citep{Hernandez-Lasheras2021EvaluatingModelling}. NR, in contrast, uses the Mercator Glorys reanalysis global product, with a 1/12$^\circ$ horizontal resolution (CMEMS GLOBAL\_REANALYSIS\_PHY\_001\_030) and has also been validated to give realistic simulations, comparing against observational data from satellites and Argo buoys. The atmospheric forcing, common for both simulations, is provided every 3 hours at $1/20^\circ$ resolution by the Spanish Meteorological Agency (AEMET) through the HIRLAM model \citep{Unden2002HighDocumentation} and the bathymetry is derived from a 1' database \citep{Smith1997GlobalSoundings}.

\begin{figure}[!ht]
  \centering
  \includegraphics[width=.98\linewidth]{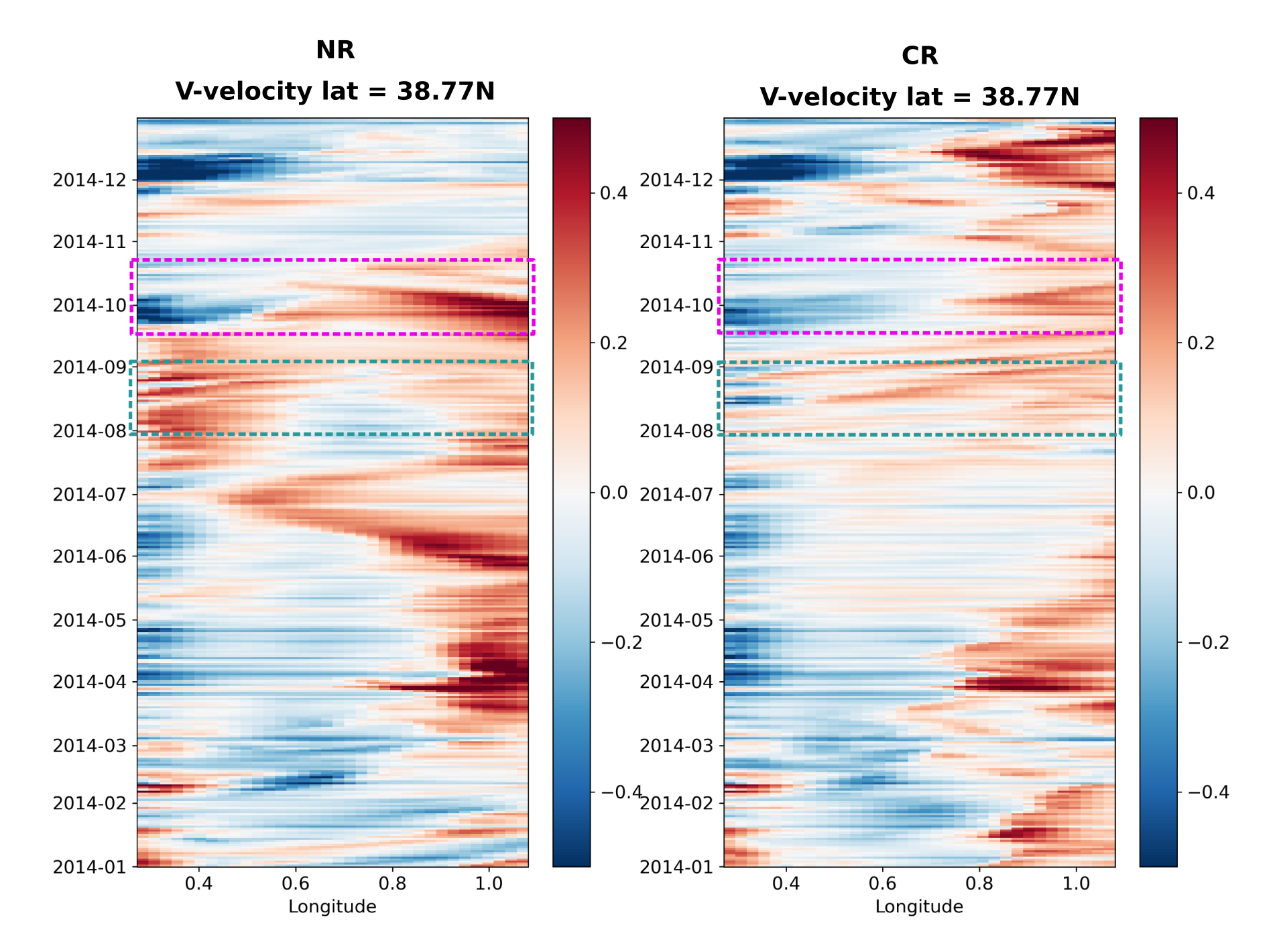}
    \caption{ Hovmoller of the meridional component of velocity during 2014 for a transect at 38.77N latitude, in the Ibiza Channel, for NR (left) and CR (right). The two periods of the OSSE are highlighted. The period coincident with the OSE is marked in pink (21 September to 20 October). In green, the second period simulated, during August 2014.}
  \label{fig:hovmoller-v-2014}
\end{figure}

\begin{figure}[!ht]
  \centering
  \includegraphics[width=.98\linewidth]{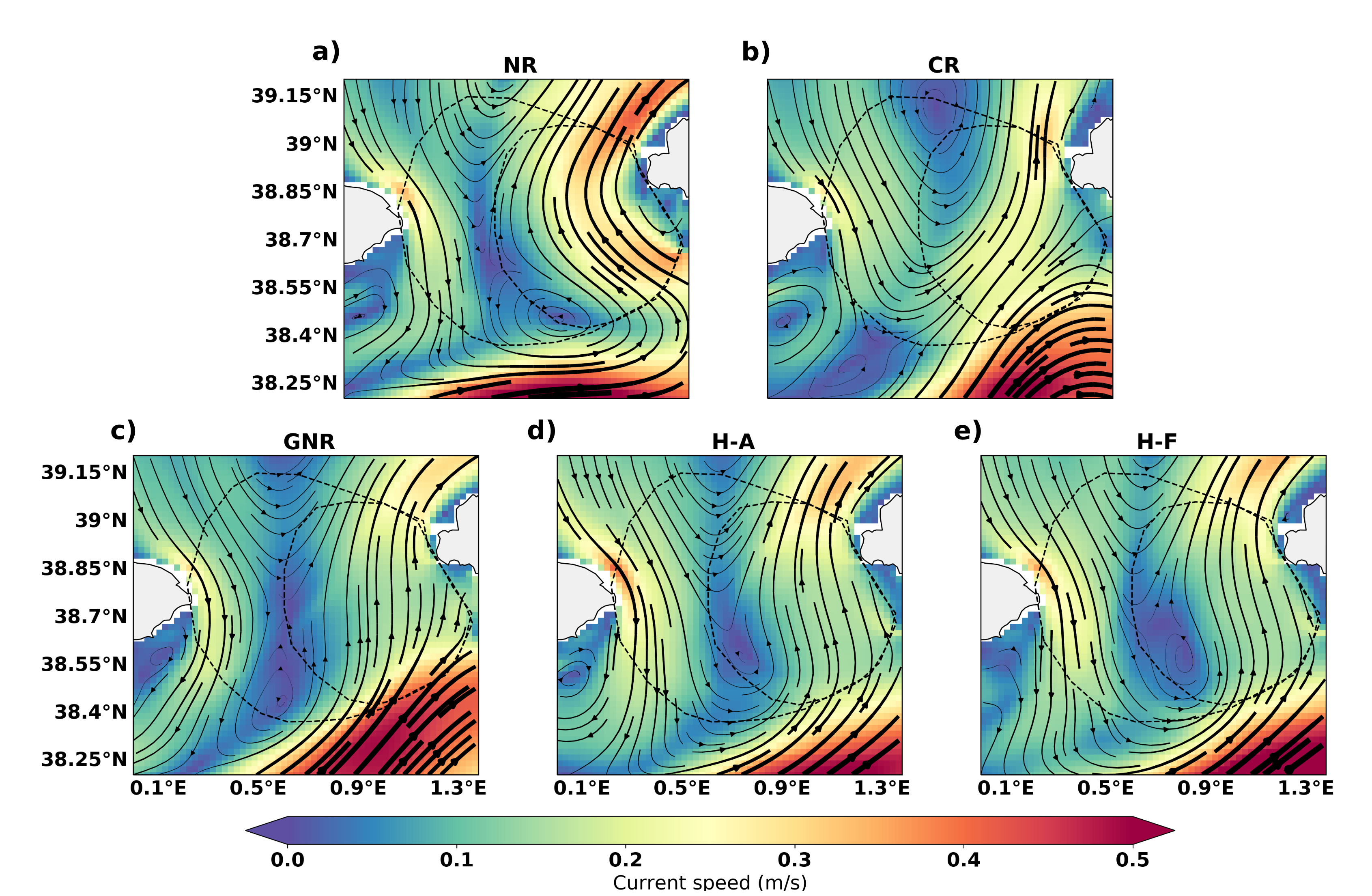}
    \caption{ Mean surface circulation during the one-month simulation (20-September to 20-October) in the Ibiza Channel  for the a)Nature Run (NR), b) Control Run (CR), c) GNR, d) H-A, e) H-F. Actual and future coverage areas are marked in dashed black lines.}
  \label{fig:mean_circulation}
\end{figure}

Both model realizations resolve the same scales, while differing in the mesoscale structures present during the experiment period, which are the two main initial requirements needed for a fraternal twin OSSE approximation. Figure \ref{fig:hovmoller-v-2014} shows the Hovmoller diagram of the meridional velocity in a transect across the Ibiza Channel (latitude 38.77$^\circ$N), where we can observe differences between both runs in the currents across the IC  during the whole year 2014. The mean circulation pattern in the Ibiza Channel between 21 September and 20 October are shown in the top two panels of Fig. \ref{fig:mean_circulation}, where we have also marked (dotted line) the coverage areas of the actual and the possible future antennas considered in this study. Both simulations present in average a southward current at the western side of the IC, while having a northward flow in the eastern part. In the case of the NR both flows are more intense than in the CR, which depicts a more intense eastward current in the southern part of the coverage area.

\begin{figure}[!ht]
  \centering
  \includegraphics[width=.98\linewidth]{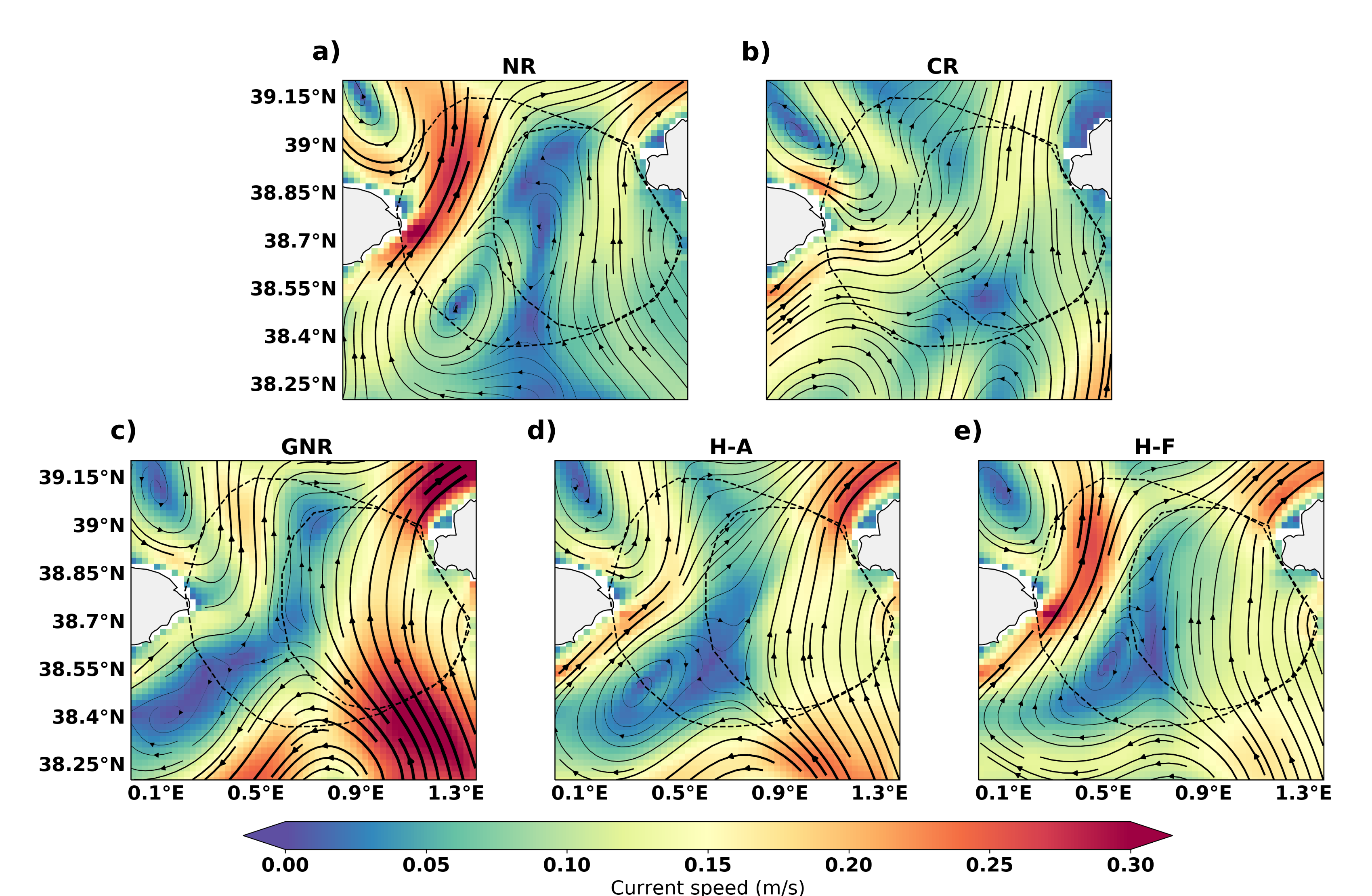}
    \caption{ Mean surface circulation during the one-month simulations (August 2014) in the Ibiza Channel for the a) Nature Run (NR), b) Control Run (CR), c) GNR, d) H-A, e) H-F. Actual and future coverage areas are marked in dashed black lines.}
  \label{fig:mean_circulation_aug}
\end{figure}

To further explore the capabilities of the new antenna under different possible circulation patterns, we selected another period during which the surface circulation from the two simulations
present larger differences in the study area. This other period is August 2014. In the CR simulation, the dynamics during August are similar to the following September-October period, with northward currents in the east side of the channel and southward in the west, as can be seen in Fig. \ref{fig:hovmoller-v-2014}. On the contrary, the NR depicts a northward current in the eastern  side of the channel and also in the west, where it is more intense. In the middle of the channel (0.62$^\circ$E-0.85$^\circ$E), there is a strip of weak northward current neither present in the CR. Top two panels of Fig. \ref{fig:mean_circulation_aug} shows the mean circulation in the region for the NR and CR.

For the two commented periods we run three data assimilative simulations, using different datasets, and we evaluate the impact comparing against the free-run CR (Table \ref{table:experiments}). We called GNR the simulation assimilating the generic data set, composed of SLA along-track, SST and Argo T-S profiles. H-A and H-F are the simulations that, additionally to this generic data-set, employ HFR simulated total observations from the actual or the future coverage area, respectively. 

The data assimilation system employed is the Local Multimodel EnOI scheme previously successfully used to assimilate satellite observations, temperature and salinity profiles from CTDs and gliders as well as surface velocities from the Ibiza Channel HFR \citep{Hernandez-Lasheras2018DenseSardinia, Hernandez-Lasheras2021EvaluatingModelling}. We here use the nudging initialization method after analysis, as it is the one employed in the WMOP operational system and it is less prone to produce discontinuities in the field which could affect the computation of FSLE, that will be later presented.


\begin{table}[!ht]
\centering{%
\resizebox{9cm}{!}{
\begin{tabular}{clc}
\hline
\multicolumn{1}{|c|}{\textbf{Simulation}}  & \multicolumn{1}{|c|}{\textbf{Assimilated observations}}  \\ \hline \hline

\multicolumn{1}{|c|}{\textbf{NR}}  & \multicolumn{1}{|c|}{\textbf{None. Pseudo-reality}}  \\ \hline \hline

\multicolumn{1}{|c|}{\textbf{CR}} & \multicolumn{1}{c|}{None}     \\ \hline
\multicolumn{1}{|c|}{\textbf{GNR}}  & \multicolumn{1}{c|}{SLA, SST, TS}        \\ \hline 
\multicolumn{1}{|c|}{\textbf{H-A}} & \multicolumn{1}{c|}{SLA, SST, TS, HFR (actual coverage)} \\ \hline
\multicolumn{1}{|c|}{\textbf{H-F}} & \multicolumn{1}{c|}{SLA, SST, TS, HFR (future coverage)} \\ \hline
\end{tabular}%
}
}
\caption{Basic description of the experiments, indicating the dataset used in the simulations. }
\label{table:experiments}
\end{table}

\subsection{OSSE Set-up: Pseudo-Observations}
\label{met:pseudo_obs}

For our experiment, the satellite and Argo pseudo observations have been extracted at the same position and time as the real observations analyzed in the OSE presented in \citep{Hernandez-Lasheras2021EvaluatingModelling}. We simulated along-track sea level anomalies (SLA) from four different altimeters (Cryosat, Jason-2, Saral Altika, and HY-2). NR fields are interpolated in space and time to each satellite observation after removing the mean dynamic topography. For SST, we emulate the SST foundation product, which does not account for the diurnal cycles, sub-sampling surface temperature fields from the NR at 8 a.m., with a 10 km resolution. The Argo profiles were sampled by interpolating the temperature and salinity fields in space and time. We added noise to every observation, which was randomly generated for each observation considering a Gaussian probability distribution with a standard deviation of the value of the error. The observation error has been considered the same as for the real observations. Table \ref{table:errors_noise} indicates the value of the representation and instrumental errors considered for the different observations. For Argo observations only the horizontal representation error is shown. Note that the total error has the expression $\sigma^{2}_{tot} = \sigma^{2}_{rep} + \sigma^{2}_{ins}$.

\begin{table}[]
\centering
\begin{tabular}{c|c|c|}
\cline{2-3}
 & \textbf{Representation} & \textbf{Instrumental} \\ \hline
\multicolumn{1}{|c|}{\textbf{SLA (m/s)}} & 0.03 & 0.02 \\ \hline
\multicolumn{1}{|c|}{\textbf{SST ($^\circ$)}} & 0.25 & 0.5 \\ \hline
\multicolumn{1}{|c|}{\textbf{Argo - T ($^\circ$)}} & 0.25 & 0.1 \\ \hline
\multicolumn{1}{|c|}{\textbf{Argo - S}} & 0.05 & 0.01 \\ \hline
\multicolumn{1}{|c|}{\textbf{HFR (m/s)}} & 0.03 & 0.02 \\ \hline
\end{tabular}
\caption{Representation and instrumental errors employed for the different observations.}
\label{table:errors_noise}
\end{table}

\begin{figure}[!ht]
  \centering
  \includegraphics[width=.98\linewidth]{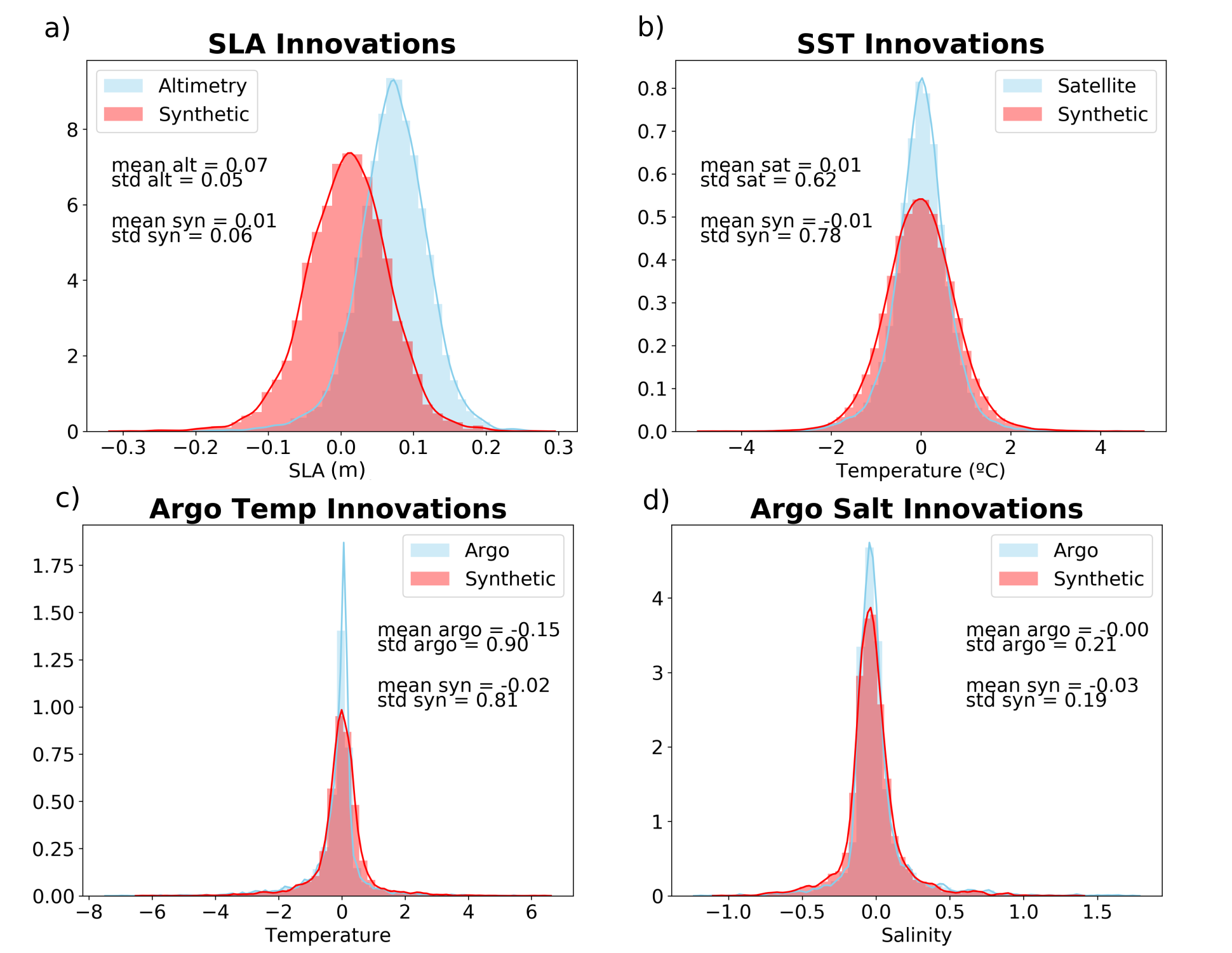}
    \caption{Histograms comparing the innovations (observation - model) for OSE (blue) and OSSE (red) for the different observing sources: a) SLA along track, b) SST, c) Argo temperature profiles, d) Argo salinity profiles. }
  \label{fig:histograms_sla-sst-argo}
\end{figure}

 The histogram of the innovations (observation - model) is shown in Fig. \ref{fig:histograms_sla-sst-argo} for the OSE (blue) and OSSE (red), whose results are also synthesized in Table \ref{table:innovation}. We can observe that all observations follow a similar distribution both for OSE and OSSE. The only discrepancy is found in the SLA, where we can observe a bias of 0.07 m in the OSE. This is a common issue linked to the mismatch between the mean dynamic topography (MDT) of the model and that of the observations. Moreover, two different simulations using different initial conditions generally also show differences in the MDT. However, we believe this does not significantly affect the assimilation of SLA, as it is not corrected during the simulation (as discussed in the previous chapter  \citep{Hernandez-Lasheras2021EvaluatingModelling}) and do not impact the geostrophic circulation. The values of the innovations standard deviation, which is directly related to the centered-root-mean-square-deviation (CRMSD), has small differences between OSE and OSSE for all observing sources. 

\begin{table}[]
\centering
\begin{tabular}{c|cc|cc|cc|cc|}
\cline{2-9}
\multicolumn{1}{l|}{}               & \multicolumn{2}{c|}{\textbf{SLA (cm)}}            & \multicolumn{2}{c|}{\textbf{SST ($^\circ$C)}}            & \multicolumn{2}{c|}{\textbf{Argo T ($^\circ$C)}}         & \multicolumn{2}{c|}{\textbf{Argo S}}              \\ \cline{2-9} 
\multicolumn{1}{r|}{\textbf{}}      & \multicolumn{1}{c|}{\textbf{mean}} & \textbf{std} & \multicolumn{1}{c|}{\textbf{mean}} & \textbf{std} & \multicolumn{1}{c|}{\textbf{mean}} & \textbf{std} & \multicolumn{1}{c|}{\textbf{mean}} & \textbf{std} \\ \hline
\multicolumn{1}{|c|}{\textbf{OSE}}  & \multicolumn{1}{c|}{0.07}          & 0.05         & \multicolumn{1}{c|}{0.01}          & 0.62         & \multicolumn{1}{c|}{-0.15}         & 0.90         & \multicolumn{1}{c|}{-0.00}         & 0.21         \\ \hline
\multicolumn{1}{|c|}{\textbf{OSSE}} & \multicolumn{1}{c|}{0.01}          & 0.06         & \multicolumn{1}{c|}{-0.01}         & 0.78         & \multicolumn{1}{c|}{-0.02}         & 0.81         & \multicolumn{1}{c|}{-0.03}         & 0.19         \\ \hline
\end{tabular}
\caption{Mean value and standard deviation of the Innovations of SLA, SST and Argo T-S for the OSE and OSSE.}
\label{table:innovation}
\end{table}

\begin{figure}[!ht]
  \centering
  \includegraphics[width=.98\linewidth]{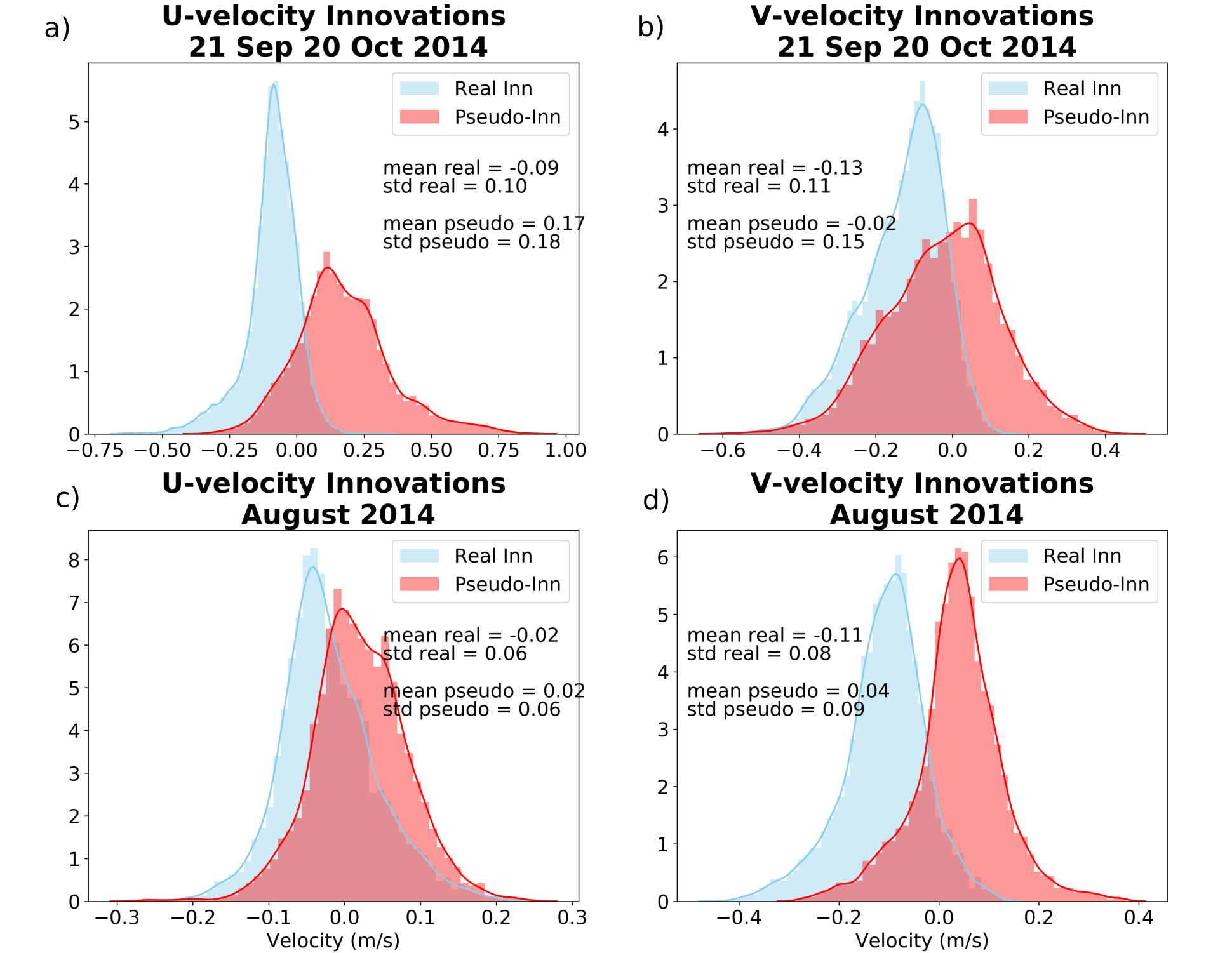}
    \caption{ Histograms comparing HFR Pseudo-obs created from NR against real observations assimilated in previous OSE: a) zonal velocity, b) meridional velocity. }
  \label{fig:histograms_hfr}
\end{figure}

For the HFR observations we have followed a slightly different approach. We have considered two polygons, one containing the actual coverage area and another considering the potential future coverage that a set of two antennas settled in the western part of the IC might provide, according to expert criteria (Fig. \ref{fig:coverage_area}). Within these areas, we have sub-sampled the daily mean velocity fields of the NR in a spatial grid of 3 km resolution, which corresponds to the HFR total (u-v) observing resolution in the area \citep{tintore2012hf}. We randomly discarded 15\% of the observations for simulating potential gaps in the antennas coverage. Again, we have introduced a Gaussian noise to the observations and considered the same instrumental and representativity error we previously used with the real data. Fig. \ref{fig:histograms_hfr} shows the histogram of the innovations for both variables of velocity during the two periods run. 

It can be observed that during the period of 21  September to 20 October, coincident with the previous OSE, the innovations present a larger discrepancy, both in mean and standard deviation, as seen in the spread of the distribution. During this period, the model tends to overestimate the meridional currents observed by the HFR. However these discrepancies are not systematic, as can be seen for the period of August, where the innovation distribution is much more similar when using real or virtual observations. While the meridional component still has a mean difference, the standard deviation is almost equal whether using real or virtual observations for both velocity components.

Overall, the statistical properties of the innovations are consistent between the OSE and the OSSE, which validates the use of the NR to generate pseudo-observations. The validation of the OSSE framework employed will be further completed, assessing the impact of the observations on the model.

\subsection{Lagrangian Analysis.}
\label{met:lagrangian}

The effects in the transport are here assessed using a Lagrangian approach. In the Lagrangian framework one has the advantage of exploiting both spatial and temporal variability of a given velocity field \citep{Hernandez-Carrasco2011HowDynamics}. In particular, we will use the finite size Lyapunov exponents (FSLE) where a set of fluid particles, initially separated a distance $\delta_0$, are transported by the flow and followed in time by integrating the equation of motion. The FSLE, denoted by $\rm{\lambda}$  (Eq. \ref{eq:fsle}), is inversely proportional to the time at which two fluid particles initially separated $\delta_0$ get separated a certain distance  $\delta_f$.  When integrated backwards, high values of FSLE are identified as limits of maximum stretching. The so-called Lagrangian Coherent Structures (LCS) are the ridges of FSLE fields, which act as transport barriers.

The integration of the trajectories is performed with OceanParcels \citep{Lange2017ParcelsAge}, using a Runge-Kutta 4 algorithm with an integration time step of 1 hour. The initial separation distance $\delta_0$ is considered equal to the model grid resolution (around 2 km). The final distance $\delta_f = 10\cdot \delta_0$, as considered optimal in other studies which explore the importance of this parameter \citep{Hernandez-Carrasco2011HowDynamics,Hernandez-Carrasco2012SeasonalOcean},\begin{equation}
    \rm{\lambda} \left ( \mathbf{r}, \mathit{t}, \rm{\delta_0}, \rm{\delta_f} \right ) \equiv  \frac{1}{\tau } log \frac{\delta_f}{\delta_0}.
\label{eq:fsle}
\end{equation}

We calculated the FSLE field in the entire domain, launching particles with an initial separation equal to the model grid size. For the one-month simulations, we computed the FSLE integrating the trajectories backwards in time during 15 days.  From the 16$^{\rm{th}}$ simulation day onward, the last 15 days' fields are used to obtain one FSLE field.


\section{Results}

\subsection{OSSE validation: Comparison with OSE.}
\label{res:osse_validation}

First, we assess the results of our experiments comparing them to the ones obtained in the companion OSEs \citep{Hernandez-Lasheras2021EvaluatingModelling}. The evaluation system will be considered as valid (i.e. it reliably estimates the impact of potential observations) when the forecast error reduction obtained in both cases (OSE and OSSE) is similar.  For this validation we analyze the period spanning from 21 September to 20 October from an Eulerian perspective, in a similar way as it was done in the previous OSE: a) For SLA, we compare for each day of simulation the model equivalents against the NR at every location of all along-track possible observations in the region; b) SST comparison between model and NR is performed at every observation point within a grid of 10 km resolution, like the one used to generate the pseudo-observations; c) Temperature and salinity fields are interpolated in time and space to the Argo float profile observations; d) For comparison against HFR data we interpolate the surface average fields to the position of the real observations. Note that in the OSSE perspective, we compare against the value of the true ocean state, represented by the NR simulation.

Table \ref{table:nCRMSD_ose_osse} shows for each observing source the CRMSD normalized with the CR for the OSE and OSSE. This metric give us an overview of the impact, without taking into account the mean error (bias), which is only significant in the case of SLA observations, as was commented previously (section \ref{met:pseudo_obs}). For SLA, SST and Argo T-S only the results of the GNR simulation are shown, as the ones obtained for H-A are almost the same when comparing against these sources. For the comparison against HFR data we show the results of GNR and H-A simulations. 
We can observe that the normalized CRMSD for GNR is slightly higher for the OSSE than for the OSE, with even an increase in the error for the v-component. On the contrary, H-A produces better results for the OSSE in the u-component. This suggests a bigger impact of adding HFR observations in the OSSE approach for this specific period.  For the rest of observing sources we can observe that the error reduction between OSE and OSSE is of the same order.  

\begin{figure}[!ht]
  \centering
  \includegraphics[width=.98\linewidth]{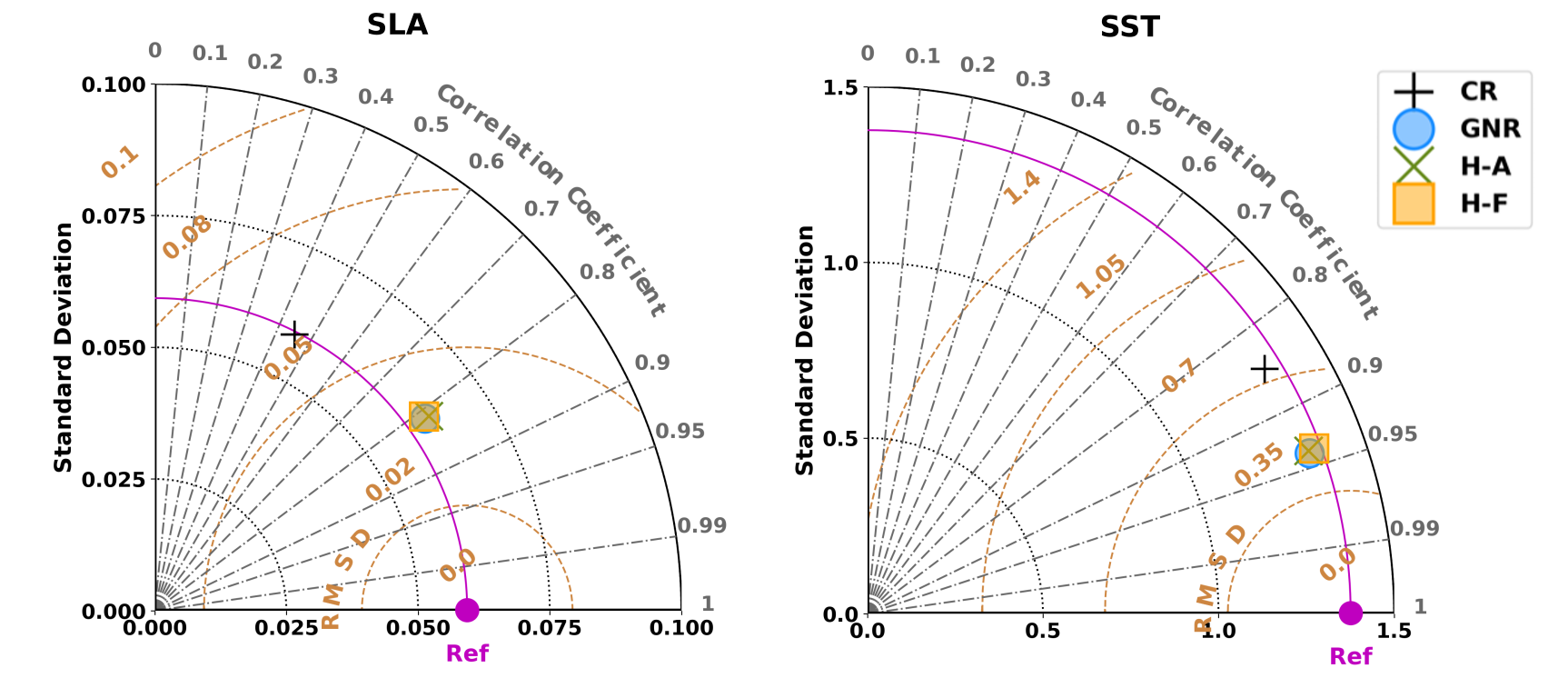}
    \caption{Taylor diagrams comparing models and NR pseudo-observations in terms of SLA (left) and SST (right) over the whole modelling domain. X and Y axis represent the standard deviations of the data. Distance from the reference point located on the X axis ( noted as Ref. in magenta) represents the centered root mean square deviation (CRMSD). Correlation between observations and model increases clockwise. Symbols represent the different simulations, as specified in the legend}
  \label{fig:taylor_SLA-SST}
\end{figure}

\begin{figure}[!ht]
  \centering
  \includegraphics[width=.98\linewidth]{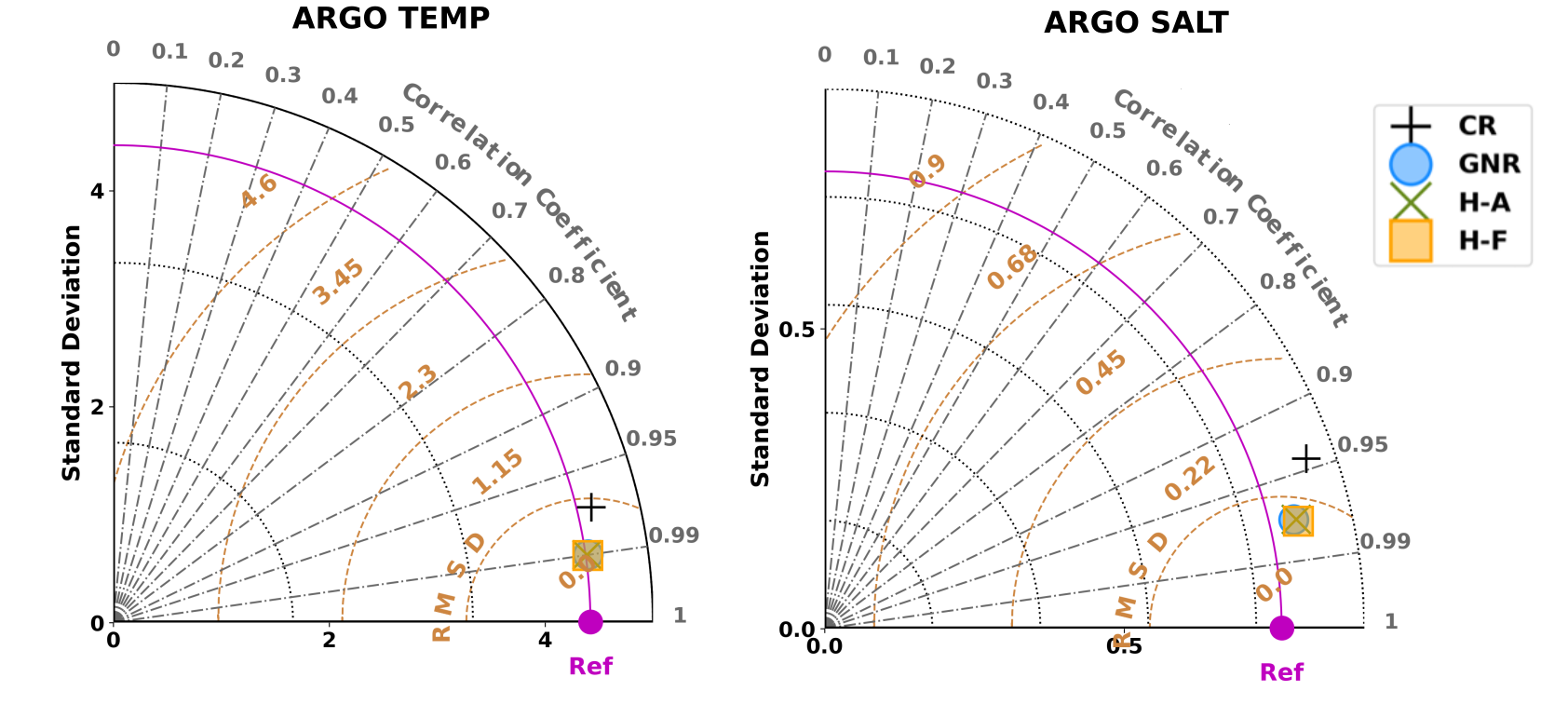}
    \caption{ Same as Fig. \ref{fig:taylor_SLA-SST} for Argo salinity and temperature }
  \label{fig:taylor_argo}
\end{figure}

A further analysis of the results is shown in the Taylor diagrams in Fig. \ref{fig:taylor_SLA-SST}. For SLA, the model error decreases around 40\% and the correlation between model and observation increases from 0.38 to 0.75 when using DA. Results are almost equal for all three simulations using DA. For SST, the comparison between model and NR is performed at every observation point within a grid of 10 km resolution, like the one used to generate the pseudo-observations. Results show how the error reduction is around 36\%, while correlation increases for all simulations from 0.87 to 0.94. These results are of the same order of magnitude as the ones obtained for real observations (shown in the previous OSE). Similarly, for the Argo T-S observations, the results obtained for the OSSE are very similar and of the same the same order as those obtained with real-world observations. The error is reduced by 41 and a 36 \%, for temperature and salinity respectively, and the correlation increases in both cases (Fig. \ref{fig:taylor_argo}).


\begin{table}[]
\centering
\begin{tabular}{c c|c|c|c|c||c|c|l}
\cline{3-8} \cline{3-8}
\textbf{}   &  & \textbf{SST} & \textbf{SLA} & \textbf{Argo-T} & \textbf{Argo-S} & \textbf{HFR-U} & \textbf{HFR-V} &  \\ \cline{1-8} \cline{1-8} 
\multicolumn{1}{|c|}{\multirow{1}{*}{\textbf{OSE}}}  & GNR & 0.69  & 0.73  & 0.66  & 0.62 & 0.80 & 0.96 &  \\ 
\multicolumn{1}{|c|}{}     & H-A  &    &   &   &    & 0.71  & 0.74  &  \\ \cline{1-8} \cline{1-8}
\multicolumn{1}{|c|}{\multirow{1}{*}{\textbf{OSSE}}} & GNR  & 0.64  & 0.60   & 0.59  & 0.64  & 0.87   & 1.06  &  \\ 
\multicolumn{1}{|c|}{}     & H-A    &    &    &     &   & 0.64    & 0.74           &  \\ \cline{1-8} \cline{1-8}
\end{tabular}
\caption{Comparative table between OSE and OSSE experiments, of the normalized RMSD against the different observing sources. SLA, SST and Argo values are only shown for GNR simulation in each case, as the values for the H-A are the same ones.}
\label{table:nCRMSD_ose_osse}
\end{table}

\subsection{Impact of the HFR system expansion: Eulerian validation}

We further assess the impact of OSSEs on the surface currents comparing the model fields in each simulation against the NR in the area of the IC. The area used for the assessment in this section, which can be seen in Fig. \ref{fig:coverage_area}, covers the entire IC, being wider than the coverage of the future HFR system. This way, we evaluate the impact of HFR DA beyond the coverage of the antennas and its effects on the transport in the region.

We evaluate the two different periods of simulation. During September-October, the meridional currents in the region are more intense, as can be seen in the Hovmoller diagram \ref{fig:hovmoller-v-2014} and so are the errors. The assimilation of generic observations only does not improve the prediction of surface currents in the region. For the u component, GNR improves the correlation with the NR from 0.15 to 0.43 but only slightly reduces the CRMSD (centered root mean square deviation). For v, both the correlation and error are slightly degraded in comparison to CR. 
The use of HFR observations additionally to the generic sources is here essential to improve the forecasting of both zonal and meridional components. The improvement obtained when assimilating observations from the future HFR system is slightly better than that obtained when only using observations from the actual coverage area. Both for H-A and H-F the correlation are higher than 0.62, and the error is reduced by more than 32\%  for the u-component. For the v-component, the correlation for both simulations also increases with respect to the NR, and the error is reduced by 15\% and 21\% for  H-A and H-F, respectively.

\begin{figure}[!ht]
  \centering
  \includegraphics[width=.98\linewidth]{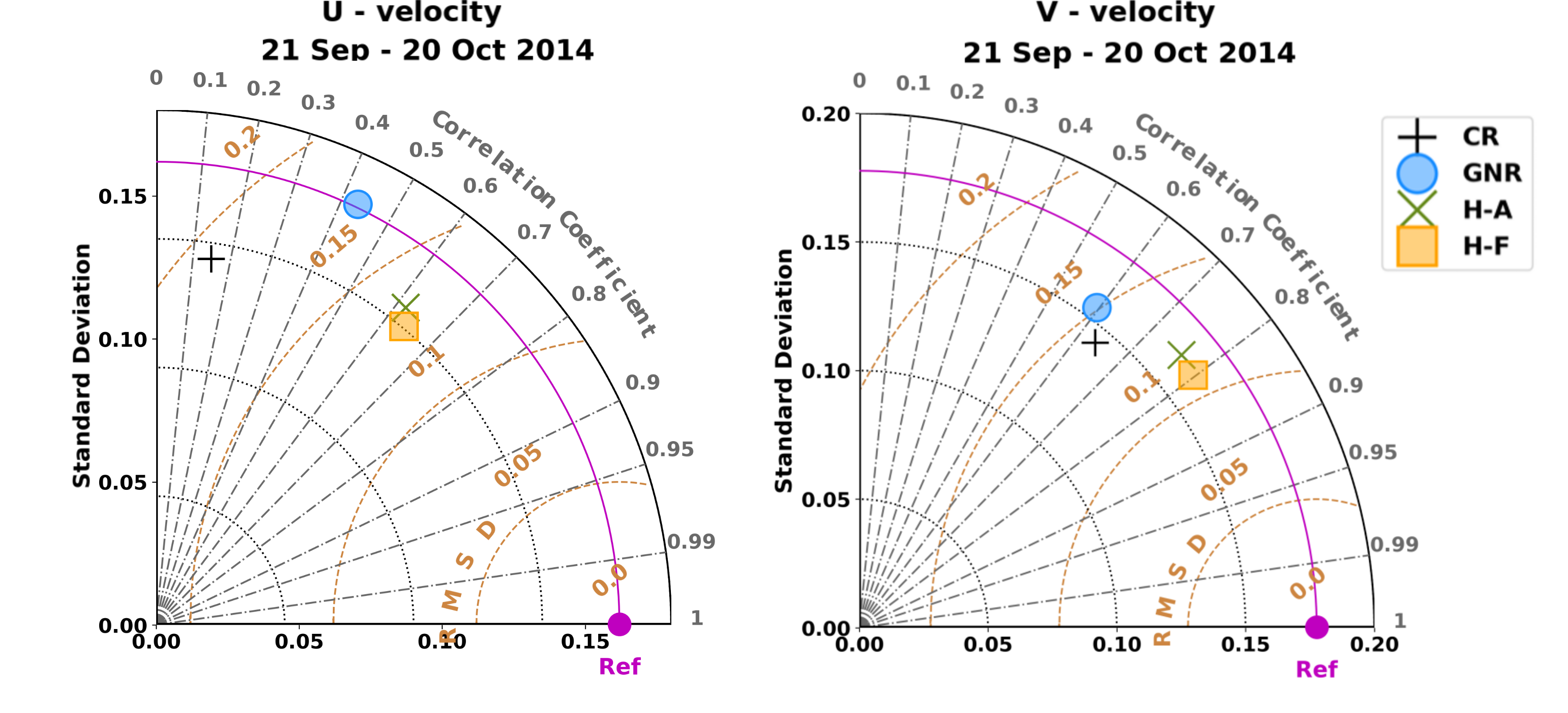}
    \caption{ Same as Fig. \ref{fig:taylor_SLA-SST} for velocities in the IC for the simulations spanning 21 September and 20 October 2014. }
  \label{fig:taylor_velocities_big_oct}
\end{figure}

During the second period  (1-30 August), the meridional currents were less intense than for the other simulation period, both in NR and CR. Furthermore, in NR, the currents in the western part of the IC were northward, contrary to the CR and the area's mean circulation pattern. This situation highlights the highly dynamical nature of the Ibiza Channel and provides an interesting study case complementary to the first study period. We aim to explore the potential impact of the HFR system extension in such situations, where the model could significantly differ from the observations. GNR degrades the forecasting of surface currents compared to CR. HFR observations are also needed to reduce the error and increase the correlation between the model simulations and the NR. When using HFR observations for the zonal velocity the correlation increases from 0.50 for the CR to 0.76 and 0.80, and the error is reduced by 23 and 29\%, for H-A and H-F respectively.

The difference between using observation only from the actual coverage area and from the future one is significant for the meridional velocity. While H-A increases the correlation from 0.39 to 0.57 and reduces the RMSD by 17\%, H-F further reduces the error by 19\% with respect to H-A, meaning a total 32\% total error reduction, with a correlation of 0.71.

\begin{figure}[!ht]
  \centering
  \includegraphics[width=.98\linewidth]{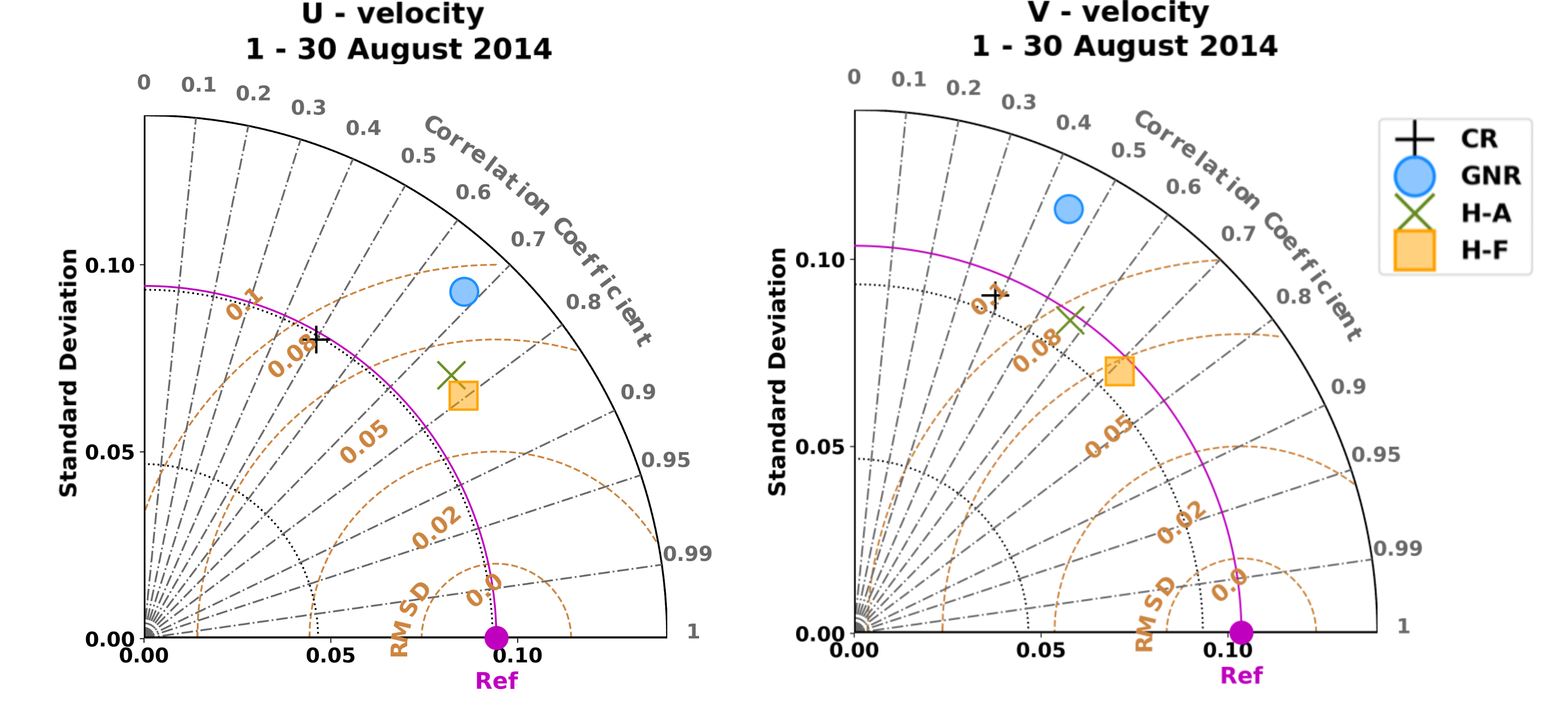}
    \caption{ Same as Fig. \ref{fig:taylor_SLA-SST} for velocities in the IC for the simulations spanning 1 and 30 August 2014. }
  \label{fig:taylor_velocities_big_aug}
\end{figure}

\subsection{Lagrangian validation}
\label{res:lagrangian}
 We focus our study on the region surrounding the IC, checking if the model can reproduce the LCS present in the NR when using DA. A qualitative analysis reveals that DA changes the LCS of the model with respect to the CR and can generate some of the structures present in the NR.

The LCS show areas of particle accumulation and barriers to transport.
To better understand how DA impacts the dynamics in the area and how the transport patterns can be modified in the IC, we  launch every 3 hours  a set of 1000 particles in 4 different regions: i.e., north, south, east, and west of the IC. The regions are selected based on a geographical situation to evaluate the zonal and meridional flow exchanges.

\begin{figure}[!ht]
  \centering
  \includegraphics[width=1.00\linewidth]{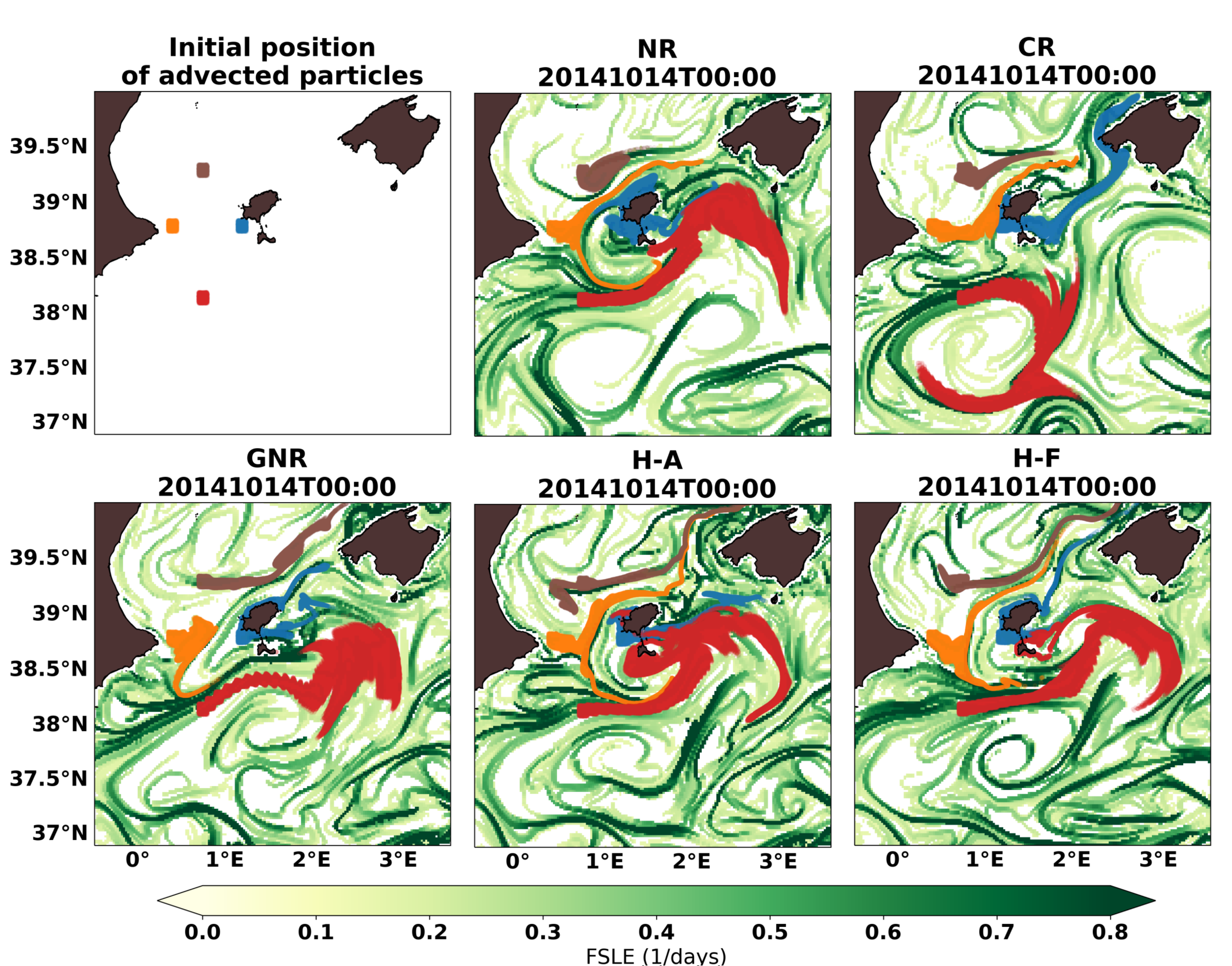}
    \caption{ Lagrangian coherent structures from FSLE calculated backwards corresponding to 16-October-2014. Particles launched every 3 hours, starting in 6 October, from 4 different areas surrounding the IC are also shown with different colors. }
  \label{fig:flse_and_particles_continuos_october}
\end{figure}

Figure \ref{fig:flse_and_particles_continuos_october} shows the FSLE field for  October 14 and the position of all the particles launched at the four sites every three hours since eight days before. The main LCS significantly differ between NR and CR in all the domain. CR shows an eddy in the southwest, centered at 1E 37.6N, that traps particles (in red) deployed at the south, while in NR, we can observe a loop-shape structure southwards. Red particles in NR move northeastwards between two LCS  that make all particles flow uniformly until they arrive east of Ibiza island, where the field is less steady and with more diffusion, driving some of the particles southwards. This behavior is reproduced in the simulations using DA. H-F is the one that better reconstructs the LCS obtained with NR velocity fields.

The LCS present in the middle of the Ibiza Channel in NR are also  well reproduced in both simulations where HFR data are assimilated. The orange particles flow eastwards until reaching this LCS, which acts as a barrier, splitting the possible track of the particles in two branches surrounding the island of Ibiza.  This situation is not reproduced in CR, where all particles flow eastwards towards Ibiza crossing to the north side after a few days.

For the blue particles deployed in the eastern side of the IC we can observe how in NR most particles spread around Ibiza. For CR, the particles are quickly advected north-eastwards reaching the north of Mallorca island after a few days, following the LCS that joins the east parts of both islands. This structure is not so intense but still present in the data-assimilative simulations. However, most of the particles still remain close to Ibiza.

Finally, the set of particles deployed at the north (brown) are more dispersed in the DA simulations, along the LCS that is formed north of Ibiza at 39.2 N approximately. This structure is present in NR, but the particles slightly move during the eight days of simulation.

\begin{figure}[!ht]
  \centering
  \includegraphics[width=1.0\linewidth]{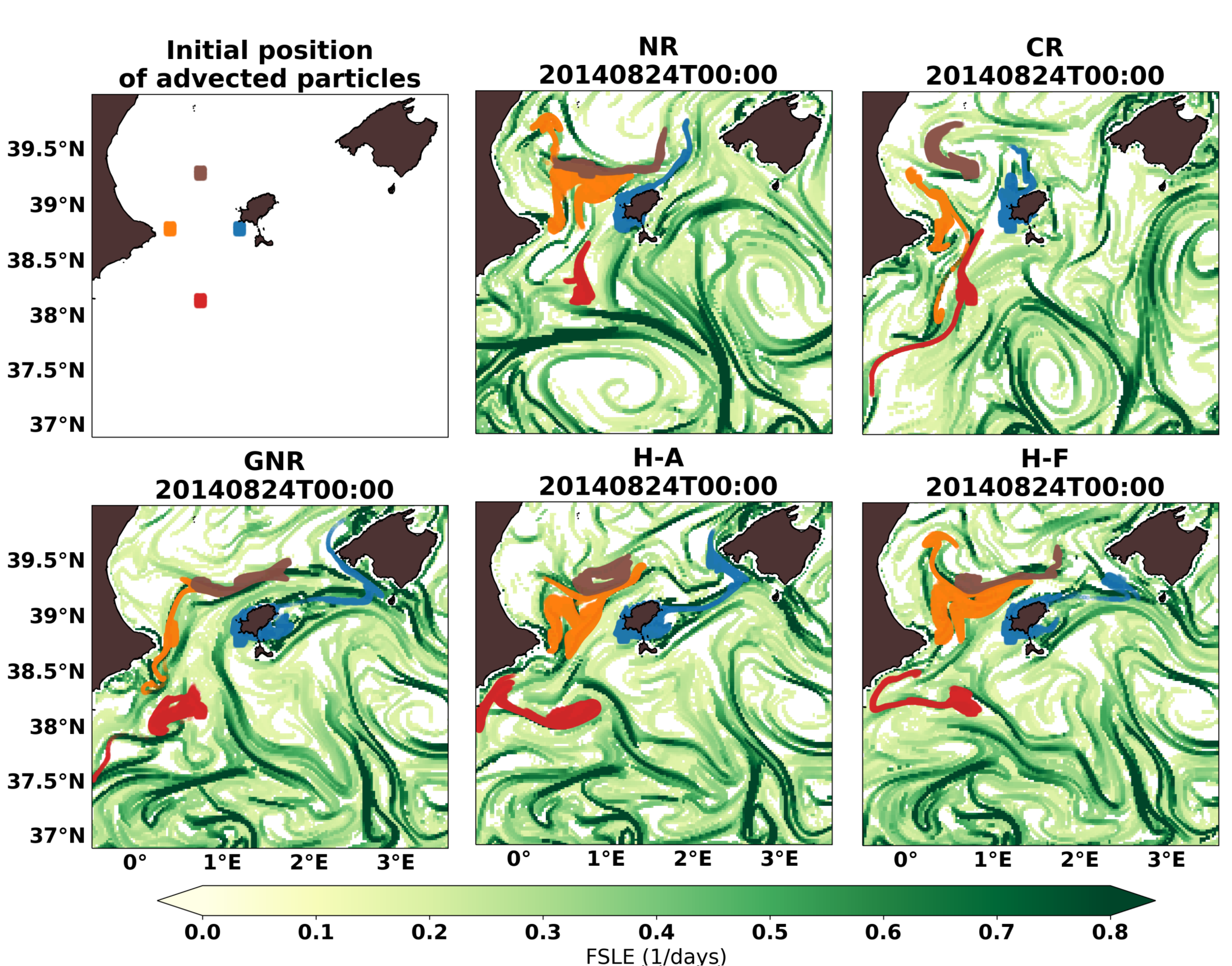}
    \caption{ Lagrangian coherent structures from FSLE calculated backwards corresponding to 24-August-2014. Particles launched every 3 hours, starting in 16 August, from 4 different areas surrounding the IC are also shown with different colors. }
  \label{fig:flse_and_particles_continuos_august}
\end{figure}

Figure \ref{fig:flse_and_particles_continuos_august} depicts the FSLE fields for 24 August, 2014 and the position of the particles, which were continuously launched every three hours since up to 8 days before. NR presents two big round shaped LCS in the south and east part of the plotted area, probably due to two respective eddies. CR also presents two big structures, but more displaced to the east. The zonal transport in the Mallorca-Ibiza channel will be restricted in the CR by a LCS that extends along the north coast of Mallorca and crossing the channel southward. On the other hand, the motion of particles is constrained meridionally in the IC, especially in the simulations with DA. In NR, the northern part of the eddy previously described would block this transport, while several structures limit it in all the DA simulations. 

The most relevant difference found, regarding the transport of particles between the different simulations, are seen in the northern and eastern sides of the channel. For NR, orange particles flow northwards, as expected by the mean current observed during this period (Fig. \ref{fig:mean_circulation_aug}), until they get blocked by a LCS, limiting the transport of brown particles at the south, that extend along the LCS in both directions. This behavior, that is not well reproduced in CR, can be reasonably well captured in H-F, where particles motions depict a very similar pattern, and also for H-A, but to a less extent. In this simulation, there is a slight displacement of some orange particles southwards at the initial stages, and we do not identify the left branch of the orange particles flowing northwards, as in NR and H-F.

\section{Discussion}

The experiments presented here apply an approach which has not been used before for the design and extension of a HFR system. The NR is validated to give realistic simulations and the innovation distribution of the pseudo-observations present a similar distribution to the real data. The results obtained in the OSSE framework in terms of error reduction and correlation increment are of the same order as the ones obtained in the previous OSE work. For surface currents, there is a significant difference between the innovations obtained in OSE and OSSE, however we have seen that this difference is not systematic and depends on the analyzed period (Figure \ref{fig:histograms_hfr}). Furthermore, when assimilating HFR, the error reduction and the increment in the correlation are also of the same order as the one achieved in the OSE.

We have evaluated the impact on the surface currents in a wider area than that covered by the antennas, taking advantage of the full ocean state knowledge provided by the OSSE framework. It draws attention that the assimilation of the generic observing sources alone cannot correct the circulation in the area in the OSSE. In the previous OSE the assimilation of generic sources led to a good improvement compared to the CR. Even though the model's response to SLA, SST and Argo observations is very similar to that obtained with real observations, the circulation seems to have a highly ageostrophic component. Therefore, the correction of SSH and density fields is not able to correct the surface currents in this region. This enhances the need of high-resolution surface current observations in coastal areas where satellite-derived geostrophic currents tend to fail \citep{Vignudelli2019SatelliteZone}. 

The extension of the HFR system seems to be useful under certain conditions. The new antennas provide a moderate effect when NR and CR simulations reproduce similar circulation patterns. However, when both simulations present different circulation patterns, especially in the western side of the channel, the availability of surface current observations over the full channel is key to improve the model circulation. This can be interpreted as real situations in which the model is unable to reproduce the dynamics observed by the HFR system. 

The observation error for the HFR has been considered the same in all the domain. However, the error could be expected to be reduced in areas covered by three antennas where a larger number of radial observations can be used to generate a total reconstructed velocities. This approach using a spatially variable error depending on the number of available radial measurements at each point has been explored by some authors \citep{Vandenbulcke2017}. The improvement of the observation error, including  correlated errors, remains an aspect to be explored in future studies. Besides, the generation of all pseudo-observations could be made more realistic by incorporating a spatially correlated noise.

We used a Lagrangian approach to evaluate the impact of data assimilation on the surface transport. The Lagrangian techniques, such as FSLE, have been increasingly used in the last years. However, the effects of computing FSLEs using a model field sequentially corrected through DA remains still poorly studied. As particles are advected, the DA simulation's discontinuities might impact the trajectories of the particles and the following computation of FSLE. Here we showed that simulations using DA behave in a similar way to those without DA. The particles tend to accumulate along LCS, which act as barriers to transport. The possible discontinuities do not seem to affect or create artifacts in the FSLE field or LCS. When comparing the FSLE fields computed for consecutive days, the transition between them is smooth. There is no significant difference when comparing the variation of LCS for two consecutive days with or without DA. Furthermore, the experiments performed here show how we can reconstruct some LCS  present in the ocean state when assimilating observations. In particular, the use of HFR data for assimilation helps to recreate the LCS present in the NR and to correct the dynamics and the transport in the region, as was demonstrated with the advection of particles.

The four different areas in the IC from which we continuously deployed particles were selected in terms of their geographical location in the IC to evaluate the zonal and meridional transports in the channel and the connectivity between the different regions. The study could also be complemented by analysing the trajectories of particles launched at different sites as the ones shown in this work. Besides, a further quantitative analysis would be desirable, even if, establishing a metric for this kind of analysis is difficult and not extended in the literature. FSLE fields should not be compared point-wise, as little differences in the position of LCS could affect the results, leading to an erroneous interpretation. Developing a valid metric to quantify LCS differences remains a future work.

OSSEs are an important tool to explore the capabilities of a future observing system design. Strictly, in scientific terms, it is always good to have as many observations as possible. However, as resources are limited, synergies between observing and modeling communities are needed for a mutual benefit, and observing systems should rigorously be designed to meet user requirements and respond to societal needs \citep{Davidson2019SynergiesObservations}. Further considerations could be taken into account for an optimal design of the observing system expansion. For instance, different locations for the antennas may be examined. For a final design, the decision should be jointly based on the scientific, and on the logistic and economic assessment, that are out of the scope of this work.


\section{Conclusions}
The objective of this study was to evaluate the impact of setting up a couple of antennas to complement the currently operating ones in the IC. We analyzed the impact of this new observing system on the transport properties through the LCS computation. The effect of data assimilation on the reconstruction of the LCS and its impact on the spreading of the advected particles has been assessed in a Lagrangian framework. A series of OSSEs assimilating HFR data along with traditional observing sources (SLA, SST, Argo) is presented here. The study complements the work from \cite{Hernandez-Lasheras2021EvaluatingModelling}, which provides a reference OSE used to validate the OSSE framework performed here. The assessment of the OSSE is consistent with that of the reference OSE and the framework is considered suitable for the design and evaluation of the impact of future observing systems. 

The impact of a potential extension of the actual HFR system in the IC has been assessed. The two new antennas would provide a full coverage of the surface currents in the IC and could help to improve the forecasting of the circulation in the region. In circumstances where the flow regime represented by the model would differ from the observed one, a diminution of up to 19\% of the error can be expected when assimilating the future system observations, compared to the present situation.

A Langrangian analysis based on FSLE revealed that the use of model outputs corrected with DA are useful for this kind of analysis and are not significantly affected by possible field discontinuities. Furthermore, the analysis showed how the assimilation of HFR observations can help to reconstruct the LCS present in the NR and constrain the circulation in the IC.


\section*{Data availability}
Data and numerical codes will be provided by the corresponding author upon request. 


\section*{Authors contribution}
JHL and BM conceptualized the OSSE. JHL, AO and IHC conceptualized the Lagrangian assessment. AS developed some of the Lagrangian tools used in this work. JHL conducted the experiments, performed the analysis and wrote the manuscript with the help of AO and BM. All authors contributed to the discussion and review of the manuscript. 

\section*{Competing interests} The authors declare that they have no conflict of interest. 

\section*{Acknowledgements}
This work was mostly performed within the SOCIB Modelling and Forecasting facility. This research has been supported by the the EU Horizon 2020 JERICO-NEXT (grant agreement no. 654410) and EuroSea (grant agreement no. 862626) projects. Alejandro Orfila acknowledges financial support from project LAMARCA (PID2021-123352OB-C31) funded by MICIN/AEI/10.13039/ 501100011033/ FEDER, UE. The authors are indebted to the Balearic Islands Coastal Observing System (SOCIB) for the HF-Radar data and modelling infrastructure. The authors are grateful to Emma Reyes for useful discussions about the Ibiza Channel HF radar setup. The authors also thank
AEMET, the Spanish Meteorological
Agency, for providing the HIRLAM atmospheric model
fields and the Copernicus Service for satellite and modelling products (https://marine.copernicus.eu/). The present research was carried out in the framework of the AEI accreditation ``Maria de Maeztu Centre of Excellence'' given to IMEDEA (CSIC-UIB) (CEX2021-001198). 

\bibliographystyle{unsrt}  
\bibliography{bibliografia}  

\begin{thebibliography}{45}
\expandafter\ifx\csname natexlab\endcsname\relax\def\natexlab#1{#1}\fi
\providecommand{\url}[1]{\texttt{#1}}
\providecommand{\href}[2]{#2}
\providecommand{\path}[1]{#1}
\providecommand{\DOIprefix}{doi:}
\providecommand{\ArXivprefix}{arXiv:}
\providecommand{\URLprefix}{URL: }
\providecommand{\Pubmedprefix}{pmid:}
\providecommand{\doi}[1]{\href{http://dx.doi.org/#1}{\path{#1}}}
\providecommand{\Pubmed}[1]{\href{pmid:#1}{\path{#1}}}
\providecommand{\bibinfo}[2]{#2}
\ifx\xfnm\relax \def\xfnm[#1]{\unskip,\space#1}\fi
\bibitem[{Aguiar et~al.(2019)Aguiar, Mourre, Juza, Reyes,
  Hern{\'{a}}ndez-Lasheras, Cutolo, Mason and
  Tintor{\'{e}}}]{Aguiar2019Multi-platformActivity}
\bibinfo{author}{Aguiar, E.}, \bibinfo{author}{Mourre, B.},
  \bibinfo{author}{Juza, M.}, \bibinfo{author}{Reyes, E.},
  \bibinfo{author}{Hern{\'{a}}ndez-Lasheras, J.}, \bibinfo{author}{Cutolo, E.},
  \bibinfo{author}{Mason, E.}, \bibinfo{author}{Tintor{\'{e}}, J.},
  \bibinfo{year}{2019}.
\newblock \bibinfo{title}{{Multi-platform model assessment in the Western
  Mediterranean Sea : impact of downscaling on the surface circulation and
  mesoscale activity}}.
\newblock \bibinfo{journal}{Ocean Dynamics.}
  \DOIprefix\doi{10.1007/s10236-019-01317-8 Multi-platform}.
\bibitem[{Atlas(1997)}]{atlas1997atmospheric}
\bibinfo{author}{Atlas, R.}, \bibinfo{year}{1997}.
\newblock \bibinfo{title}{Atmospheric observations and experiments to assess
  their usefulness in data assimilation (gtspecial issueltdata assimilation in
  meteology and oceanography: theory and practice)}.
\newblock \bibinfo{journal}{Journal of the Meteorological Society of Japan.
  Ser. II} \bibinfo{volume}{75}, \bibinfo{pages}{111--130}.
\bibitem[{Ballabrera-Poy et~al.(2007)Ballabrera-Poy, Hackert, Murtugudde and
  Busalacchi}]{ballabrera2007observing}
\bibinfo{author}{Ballabrera-Poy, J.}, \bibinfo{author}{Hackert, E.},
  \bibinfo{author}{Murtugudde, R.}, \bibinfo{author}{Busalacchi, A.J.},
  \bibinfo{year}{2007}.
\newblock \bibinfo{title}{An observing system simulation experiment for an
  optimal moored instrument array in the tropical indian ocean}.
\newblock \bibinfo{journal}{Journal of Climate} \bibinfo{volume}{20},
  \bibinfo{pages}{3284--3299}.
\bibitem[{Lorente et~al.(2022)Lorente, Aguiar, Bendoni, Berta, Brandini,
  Caceres-Euse, Capodici, Cianelli, Ciraolo, Corgnati, Dadic, Doronzo, Drago,
  Dumas, Falco, Fattorini, Gauci, Gomez, Griffa, Guerin, Hernandez-Carrasco,
  Hernandez-Lasheras, Licer, Magaldi, Mantovani, Mihanovic, Molcard, Mourre,
  Orfila, Revelard, Reyes, Sanchez, Saviano, Sciascia, Taddei, Tintore, Toledo,
  Ursella, Uttieri, Vilibic, Zambianchi and Cardin}]{Lorente2022}
\bibinfo{author}{Lorente, P.}, \bibinfo{author}{Aguiar, E.},
  \bibinfo{author}{Bendoni, M.}, \bibinfo{author}{Berta, M.},
  \bibinfo{author}{Brandini, C.}, \bibinfo{author}{Caceres-Euse, A.},
  \bibinfo{author}{Capodici, F.}, \bibinfo{author}{Cianelli, D.},
  \bibinfo{author}{Ciraolo, G.}, \bibinfo{author}{Corgnati, L.},
  \bibinfo{author}{Dadic, V.}, \bibinfo{author}{Doronzo, B.},
  \bibinfo{author}{Drago, A.}, \bibinfo{author}{Dumas, D.},
  \bibinfo{author}{Falco, P.}, \bibinfo{author}{Fattorini, M.},
  \bibinfo{author}{Gauci, A.}, \bibinfo{author}{Gomez, R.},
  \bibinfo{author}{Griffa, A.}, \bibinfo{author}{Guerin, C.A.},
  \bibinfo{author}{Hernandez-Carrasco, I.},
  \bibinfo{author}{Hernandez-Lasheras, J.}, \bibinfo{author}{Licer, M.},
  \bibinfo{author}{Magaldi, M.G.}, \bibinfo{author}{Mantovani, C.},
  \bibinfo{author}{Mihanovic, H.}, \bibinfo{author}{Molcard, A.},
  \bibinfo{author}{Mourre, B.}, \bibinfo{author}{Orfila, A.},
  \bibinfo{author}{Revelard, A.}, \bibinfo{author}{Reyes, E.},
  \bibinfo{author}{Sanchez, J.}, \bibinfo{author}{Saviano, S.},
  \bibinfo{author}{Sciascia, R.}, \bibinfo{author}{Taddei, S.},
  \bibinfo{author}{Tintore, J.}, \bibinfo{author}{Toledo, Y.},
  \bibinfo{author}{Ursella, L.}, \bibinfo{author}{Uttieri, M.},
  \bibinfo{author}{Vilibic, I.}, \bibinfo{author}{Zambianchi, E.},
  \bibinfo{author}{Cardin, V.}, \bibinfo{year}{2022}.
\newblock \bibinfo{title}{Coastal high-frequency radars in the mediterranean -
  part 1: Status of operations and a framework for future development}.
\newblock \bibinfo{journal}{Ocean Science} \bibinfo{volume}{18},
  \bibinfo{pages}{761 – 795}.
\newblock \DOIprefix\doi{10.5194/os-18-761-2022}.
\bibitem[{Reyes et~al.(2022)Reyes, Aguiar, Bendoni, Berta, Brandini,
  Caceres-Euse, Capodici, Cardin, Cianelli, Ciraolo, Corgnati, Dadic, Doronzo,
  Drago, Dumas, Falco, Fattorini, Fernandes, Gauci, Gomez, Griffa, Guerin,
  Hernandez-Carrasco, Hernandez-Lasheras, Licer, Lorente, Magaldi, Mantovani,
  Mihanovic, Molcard, Mourre, Revelard, Reyes-Suarez, Saviano, Sciascia,
  Taddei, Tintore, Toledo, Uttieri, Vilibic, Zambianchi and Orfila}]{Reyes2022}
\bibinfo{author}{Reyes, E.}, \bibinfo{author}{Aguiar, E.},
  \bibinfo{author}{Bendoni, M.}, \bibinfo{author}{Berta, M.},
  \bibinfo{author}{Brandini, C.}, \bibinfo{author}{Caceres-Euse, A.},
  \bibinfo{author}{Capodici, F.}, \bibinfo{author}{Cardin, V.},
  \bibinfo{author}{Cianelli, D.}, \bibinfo{author}{Ciraolo, G.},
  \bibinfo{author}{Corgnati, L.}, \bibinfo{author}{Dadic, V.},
  \bibinfo{author}{Doronzo, B.}, \bibinfo{author}{Drago, A.},
  \bibinfo{author}{Dumas, D.}, \bibinfo{author}{Falco, P.},
  \bibinfo{author}{Fattorini, M.}, \bibinfo{author}{Fernandes, M.J.},
  \bibinfo{author}{Gauci, A.}, \bibinfo{author}{Gomez, R.},
  \bibinfo{author}{Griffa, A.}, \bibinfo{author}{Guerin, C.A.},
  \bibinfo{author}{Hernandez-Carrasco, I.},
  \bibinfo{author}{Hernandez-Lasheras, J.}, \bibinfo{author}{Licer, M.},
  \bibinfo{author}{Lorente, P.}, \bibinfo{author}{Magaldi, M.G.},
  \bibinfo{author}{Mantovani, C.}, \bibinfo{author}{Mihanovic, H.},
  \bibinfo{author}{Molcard, A.}, \bibinfo{author}{Mourre, B.},
  \bibinfo{author}{Revelard, A.}, \bibinfo{author}{Reyes-Suarez, C.},
  \bibinfo{author}{Saviano, S.}, \bibinfo{author}{Sciascia, R.},
  \bibinfo{author}{Taddei, S.}, \bibinfo{author}{Tintore, J.},
  \bibinfo{author}{Toledo, Y.}, \bibinfo{author}{Uttieri, M.},
  \bibinfo{author}{Vilibic, I.}, \bibinfo{author}{Zambianchi, E.},
  \bibinfo{author}{Orfila, A.}, \bibinfo{year}{2022}.
\newblock \bibinfo{title}{Coastal high-frequency radars in the mediterranean -
  part 2: Applications in support of science priorities and societal needs}.
\newblock \bibinfo{journal}{Ocean Science} \bibinfo{volume}{18},
  \bibinfo{pages}{797 – 837}.
\newblock \DOIprefix\doi{10.5194/os-18-797-2022}.
\bibitem[{Benkiran et~al.(2021)Benkiran, Ruggiero, Greiner, Le~Traon,
  R{\'{e}}my, Lellouche, Bourdall{\'{e}}-Badie, Drillet and
  Tchonang}]{Benkiran2021AssessingMethods}
\bibinfo{author}{Benkiran, M.}, \bibinfo{author}{Ruggiero, G.},
  \bibinfo{author}{Greiner, E.}, \bibinfo{author}{Le~Traon, P.Y.},
  \bibinfo{author}{R{\'{e}}my, E.}, \bibinfo{author}{Lellouche, J.M.},
  \bibinfo{author}{Bourdall{\'{e}}-Badie, R.}, \bibinfo{author}{Drillet, Y.},
  \bibinfo{author}{Tchonang, B.}, \bibinfo{year}{2021}.
\newblock \bibinfo{title}{{Assessing the Impact of the Assimilation of SWOT
  Observations in a Global High-Resolution Analysis and Forecasting System Part
  1: Methods}}.
\newblock \bibinfo{journal}{Frontiers in Marine Science} \bibinfo{volume}{8},
  \bibinfo{pages}{1--19}.
\newblock \DOIprefix\doi{10.3389/fmars.2021.691955}.
\bibitem[{Davidson et~al.(2019)Davidson, Alvera-Azc{\'{a}}rate, Barth,
  Brassington, Chassignet, Clementi, De~Mey-Fr{\'{e}}maux, Divakaran, Harris,
  Hernandez, Hogan, Hole, Holt, Liu, Lu, Lorente, Maksymczuk, Martin, Mehra,
  Melsom, Mo, Moore, Oddo, Pascual, Pequignet, Kourafalou, Ryan, Siddorn,
  Smith, Spindler, Spindler, Stanev, Staneva, Storto, Tanajura, Vinayachandran,
  Wan, Wang, Zhang, Zhu and Zu}]{Davidson2019SynergiesObservations}
\bibinfo{author}{Davidson, F.}, \bibinfo{author}{Alvera-Azc{\'{a}}rate, A.},
  \bibinfo{author}{Barth, A.}, \bibinfo{author}{Brassington, G.B.},
  \bibinfo{author}{Chassignet, E.P.}, \bibinfo{author}{Clementi, E.},
  \bibinfo{author}{De~Mey-Fr{\'{e}}maux, P.}, \bibinfo{author}{Divakaran, P.},
  \bibinfo{author}{Harris, C.}, \bibinfo{author}{Hernandez, F.},
  \bibinfo{author}{Hogan, P.}, \bibinfo{author}{Hole, L.R.},
  \bibinfo{author}{Holt, J.}, \bibinfo{author}{Liu, G.}, \bibinfo{author}{Lu,
  Y.}, \bibinfo{author}{Lorente, P.}, \bibinfo{author}{Maksymczuk, J.},
  \bibinfo{author}{Martin, M.}, \bibinfo{author}{Mehra, A.},
  \bibinfo{author}{Melsom, A.}, \bibinfo{author}{Mo, H.},
  \bibinfo{author}{Moore, A.}, \bibinfo{author}{Oddo, P.},
  \bibinfo{author}{Pascual, A.}, \bibinfo{author}{Pequignet, A.C.},
  \bibinfo{author}{Kourafalou, V.}, \bibinfo{author}{Ryan, A.},
  \bibinfo{author}{Siddorn, J.}, \bibinfo{author}{Smith, G.},
  \bibinfo{author}{Spindler, D.}, \bibinfo{author}{Spindler, T.},
  \bibinfo{author}{Stanev, E.V.}, \bibinfo{author}{Staneva, J.},
  \bibinfo{author}{Storto, A.}, \bibinfo{author}{Tanajura, C.},
  \bibinfo{author}{Vinayachandran, P.N.}, \bibinfo{author}{Wan, L.},
  \bibinfo{author}{Wang, H.}, \bibinfo{author}{Zhang, Y.},
  \bibinfo{author}{Zhu, X.}, \bibinfo{author}{Zu, Z.}, \bibinfo{year}{2019}.
\newblock \bibinfo{title}{{Synergies in operational oceanography: The intrinsic
  need for sustained ocean observations}}.
\newblock \bibinfo{journal}{Frontiers in Marine Science} \bibinfo{volume}{6},
  \bibinfo{pages}{1--18}.
\newblock \DOIprefix\doi{10.3389/fmars.2019.00450}.
\bibitem[{deYoung et~al.(2019)deYoung, Visbeck, Filho, Baringer, Black, Buch,
  Canonico, Coelho, Duha, Edwards, Fischer, Fritz, Ketelhake, Muelbert,
  Monteiro, Nolan, O'Rourke, Ott, Le~Traon, Pouliquen, PInto, Tanhua, Velho and
  Willis}]{deYoung2019An2030}
\bibinfo{author}{deYoung, B.}, \bibinfo{author}{Visbeck, M.},
  \bibinfo{author}{Filho, M.C.}, \bibinfo{author}{Baringer, M.O.},
  \bibinfo{author}{Black, C.A.}, \bibinfo{author}{Buch, E.},
  \bibinfo{author}{Canonico, G.}, \bibinfo{author}{Coelho, P.},
  \bibinfo{author}{Duha, J.T.}, \bibinfo{author}{Edwards, M.},
  \bibinfo{author}{Fischer, A.S.}, \bibinfo{author}{Fritz, J.S.},
  \bibinfo{author}{Ketelhake, S.}, \bibinfo{author}{Muelbert, J.H.},
  \bibinfo{author}{Monteiro, P.}, \bibinfo{author}{Nolan, G.},
  \bibinfo{author}{O'Rourke, E.}, \bibinfo{author}{Ott, M.},
  \bibinfo{author}{Le~Traon, P.Y.}, \bibinfo{author}{Pouliquen, S.},
  \bibinfo{author}{PInto, I.S.}, \bibinfo{author}{Tanhua, T.},
  \bibinfo{author}{Velho, F.}, \bibinfo{author}{Willis, Z.},
  \bibinfo{year}{2019}.
\newblock \bibinfo{title}{{An integrated all-Atlantic ocean observing system in
  2030}}.
\newblock \bibinfo{journal}{Frontiers in Marine Science} \bibinfo{volume}{6},
  \bibinfo{pages}{1--22}.
\newblock \DOIprefix\doi{10.3389/fmars.2019.00428}.
\bibitem[{D{\'\i}az-Barroso et~al.(2022)D{\'\i}az-Barroso,
  Hern{\'a}ndez-Carrasco, Orfila, Reglero, Balb{\'\i}n, Hidalgo, Tintor{\'e},
  Alemany and {\'A}lvarez-Berastegui}]{diaz2022singularities}
\bibinfo{author}{D{\'\i}az-Barroso, L.},
  \bibinfo{author}{Hern{\'a}ndez-Carrasco, I.}, \bibinfo{author}{Orfila, A.},
  \bibinfo{author}{Reglero, P.}, \bibinfo{author}{Balb{\'\i}n, R.},
  \bibinfo{author}{Hidalgo, M.}, \bibinfo{author}{Tintor{\'e}, J.},
  \bibinfo{author}{Alemany, F.}, \bibinfo{author}{{\'A}lvarez-Berastegui, D.},
  \bibinfo{year}{2022}.
\newblock \bibinfo{title}{Singularities of surface mixing activity in the
  western mediterranean influence bluefin tuna larval habitats}.
\newblock \bibinfo{journal}{Marine Ecology Progress Series}
  \bibinfo{volume}{685}, \bibinfo{pages}{69--84}.
\bibitem[{d'Ovidio et~al.(2004)d'Ovidio, Fern{\'{a}}ndez,
  Hern{\'{a}}ndez-Garc{\'{i}}a and L{\'{o}}pez}]{dOvidio2004MixingExponents}
\bibinfo{author}{d'Ovidio, F.}, \bibinfo{author}{Fern{\'{a}}ndez, V.},
  \bibinfo{author}{Hern{\'{a}}ndez-Garc{\'{i}}a, E.},
  \bibinfo{author}{L{\'{o}}pez, C.}, \bibinfo{year}{2004}.
\newblock \bibinfo{title}{{Mixing structures in the Mediterranean Sea from
  finite-size Lyapunov exponents}}.
\newblock \bibinfo{journal}{Geophysical Research Letters} \bibinfo{volume}{31},
  \bibinfo{pages}{1--4}.
\newblock \DOIprefix\doi{10.1029/2004GL020328}.
\bibitem[{Farcy et~al.(2019)Farcy, Durand, Charria, Painting, Tamminem,
  Collingridge, Gr{\'{e}}mare, Delauney and
  Puillat}]{Farcy2019TowardInfrastructure}
\bibinfo{author}{Farcy, P.}, \bibinfo{author}{Durand, D.},
  \bibinfo{author}{Charria, G.}, \bibinfo{author}{Painting, S.J.},
  \bibinfo{author}{Tamminem, T.}, \bibinfo{author}{Collingridge, K.},
  \bibinfo{author}{Gr{\'{e}}mare, A.J.}, \bibinfo{author}{Delauney, L.},
  \bibinfo{author}{Puillat, I.}, \bibinfo{year}{2019}.
\newblock \bibinfo{title}{{Toward a European coastal observing network to
  provide better answers to science and to societal challenges; the JERICO
  research infrastructure}}.
\newblock \bibinfo{journal}{Frontiers in Marine Science} \bibinfo{volume}{6},
  \bibinfo{pages}{1--13}.
\newblock \DOIprefix\doi{10.3389/fmars.2019.00529}.
\bibitem[{Gasparin et~al.(2019)Gasparin, Guinehut, Mao, Mirouze, R{\'{e}}my,
  King, Hamon, Reid, Storto, Le~Traon, Martin and
  Masina}]{Gasparin2019RequirementsExperiments}
\bibinfo{author}{Gasparin, F.}, \bibinfo{author}{Guinehut, S.},
  \bibinfo{author}{Mao, C.}, \bibinfo{author}{Mirouze, I.},
  \bibinfo{author}{R{\'{e}}my, E.}, \bibinfo{author}{King, R.R.},
  \bibinfo{author}{Hamon, M.}, \bibinfo{author}{Reid, R.},
  \bibinfo{author}{Storto, A.}, \bibinfo{author}{Le~Traon, P.Y.},
  \bibinfo{author}{Martin, M.J.}, \bibinfo{author}{Masina, S.},
  \bibinfo{year}{2019}.
\newblock \bibinfo{title}{{Requirements for an Integrated in situ Atlantic
  Ocean Observing System From Coordinated Observing System Simulation
  Experiments}}.
\newblock \bibinfo{journal}{Frontiers in Marine Science} \bibinfo{volume}{6}.
\newblock \DOIprefix\doi{10.3389/fmars.2019.00083}.
\bibitem[{Guinehut et~al.(2004)Guinehut, Le~Traon, Larnicol and
  Philipps}]{Guinehut2004CombiningObservations}
\bibinfo{author}{Guinehut, S.}, \bibinfo{author}{Le~Traon, P.Y.},
  \bibinfo{author}{Larnicol, G.}, \bibinfo{author}{Philipps, S.},
  \bibinfo{year}{2004}.
\newblock \bibinfo{title}{{Combining Argo and remote-sensing data to estimate
  the ocean three-dimensional temperature fields - A first approach based on
  simulated observations}}.
\newblock \bibinfo{journal}{Journal of Marine Systems} \bibinfo{volume}{46},
  \bibinfo{pages}{85--98}.
\newblock \DOIprefix\doi{10.1016/j.jmarsys.2003.11.022}.
\bibitem[{Halliwell et~al.(2014)Halliwell, Srinivasan, Kourafalou, Yang,
  Willey, Le~H{\'{e}}naff and Atlas}]{Halliwell2014RigorousMexico}
\bibinfo{author}{Halliwell, J.R.}, \bibinfo{author}{Srinivasan, A.},
  \bibinfo{author}{Kourafalou, V.}, \bibinfo{author}{Yang, H.},
  \bibinfo{author}{Willey, D.}, \bibinfo{author}{Le~H{\'{e}}naff, M.},
  \bibinfo{author}{Atlas, R.}, \bibinfo{year}{2014}.
\newblock \bibinfo{title}{{Rigorous evaluation of a fraternal twin ocean OSSE
  system for the open Gulf of Mexico}}.
\newblock \bibinfo{journal}{Journal of Atmospheric and Oceanic Technology}
  \bibinfo{volume}{31}, \bibinfo{pages}{105--130}.
\newblock \DOIprefix\doi{10.1175/JTECH-D-13-00011.1}.
\bibitem[{Hern{\'{a}}ndez-Carrasco et~al.(2011)Hern{\'{a}}ndez-Carrasco,
  L{\'{o}}pez, Hern{\'{a}}ndez-Garc{\'{i}}a and
  Turiel}]{Hernandez-Carrasco2011HowDynamics}
\bibinfo{author}{Hern{\'{a}}ndez-Carrasco, I.}, \bibinfo{author}{L{\'{o}}pez,
  C.}, \bibinfo{author}{Hern{\'{a}}ndez-Garc{\'{i}}a, E.},
  \bibinfo{author}{Turiel, A.}, \bibinfo{year}{2011}.
\newblock \bibinfo{title}{{How reliable are finite-size Lyapunov exponents for
  the assessment of ocean dynamics?}}
\newblock \bibinfo{journal}{Ocean Modelling} \bibinfo{volume}{36},
  \bibinfo{pages}{208--218}.
\newblock \DOIprefix\doi{10.1016/j.ocemod.2010.12.006}.
\bibitem[{Hernandez-Carrasco et~al.(2012)Hernandez-Carrasco, L{\'{o}}pez,
  Hernandez-Garc{\'{i}}a and Turiel}]{Hernandez-Carrasco2012SeasonalOcean}
\bibinfo{author}{Hernandez-Carrasco, I.}, \bibinfo{author}{L{\'{o}}pez, C.},
  \bibinfo{author}{Hernandez-Garc{\'{i}}a, E.}, \bibinfo{author}{Turiel, A.},
  \bibinfo{year}{2012}.
\newblock \bibinfo{title}{{Seasonal and regional characterization of horizontal
  stirring in the global ocean}}.
\newblock \bibinfo{journal}{Journal of Geophysical Research: Oceans}
  \bibinfo{volume}{117}, \bibinfo{pages}{1--12}.
\newblock \DOIprefix\doi{10.1029/2012JC008222}.
\bibitem[{Hern{\'{a}}ndez-Carrasco et~al.(2018)Hern{\'{a}}ndez-Carrasco,
  Orfila, Rossi and Gar{\c{c}}on}]{Hernandez-Carrasco2018EffectSeas}
\bibinfo{author}{Hern{\'{a}}ndez-Carrasco, I.}, \bibinfo{author}{Orfila, A.},
  \bibinfo{author}{Rossi, V.}, \bibinfo{author}{Gar{\c{c}}on, V.},
  \bibinfo{year}{2018}.
\newblock \bibinfo{title}{{Effect of small scale transport processes on
  phytoplankton distribution in coastal seas}}.
\newblock \bibinfo{journal}{Scientific Reports} \bibinfo{volume}{8},
  \bibinfo{pages}{1--13}.
\newblock \DOIprefix\doi{10.1038/s41598-018-26857-9}.
\bibitem[{Hernandez-Lasheras et~al.(2021)Hernandez-Lasheras, Mourre, Orfila,
  Santana, Reyes and Tintor{\'{e}}}]{Hernandez-Lasheras2021EvaluatingModelling}
\bibinfo{author}{Hernandez-Lasheras, J.}, \bibinfo{author}{Mourre, B.},
  \bibinfo{author}{Orfila, A.}, \bibinfo{author}{Santana, A.},
  \bibinfo{author}{Reyes, E.}, \bibinfo{author}{Tintor{\'{e}}, J.},
  \bibinfo{year}{2021}.
\newblock \bibinfo{title}{{Evaluating High-Frequency radar data assimilation
  impact in coastal ocean operational modelling}}.
\newblock \bibinfo{journal}{Ocean Science Discussions} ,
  \bibinfo{pages}{1--29}\DOIprefix\doi{10.5194/os-2021-34}.
\bibitem[{Aydo{\u{g}}du et~al.(2018)Aydo{\u{g}}du, Hoar, Vukicevic, Anderson,
  Pinardi, Karspeck, Hendricks, Collins, Macchia and
  {\"O}zsoy}]{aydougdu2018osse}
\bibinfo{author}{Aydo{\u{g}}du, A.}, \bibinfo{author}{Hoar, T.J.},
  \bibinfo{author}{Vukicevic, T.}, \bibinfo{author}{Anderson, J.L.},
  \bibinfo{author}{Pinardi, N.}, \bibinfo{author}{Karspeck, A.},
  \bibinfo{author}{Hendricks, J.}, \bibinfo{author}{Collins, N.},
  \bibinfo{author}{Macchia, F.}, \bibinfo{author}{{\"O}zsoy, E.},
  \bibinfo{year}{2018}.
\newblock \bibinfo{title}{Osse for a sustainable marine observing network in
  the sea of marmara}.
\newblock \bibinfo{journal}{Nonlinear Processes in Geophysics}
  \bibinfo{volume}{25}, \bibinfo{pages}{537--551}.
\bibitem[{Aydo{\u{g}}du et~al.(2016)Aydo{\u{g}}du, Pinardi, Pistoia,
  Martinelli, Belardinelli and Sparnocchia}]{aydougdu2016assimilation}
\bibinfo{author}{Aydo{\u{g}}du, A.}, \bibinfo{author}{Pinardi, N.},
  \bibinfo{author}{Pistoia, J.}, \bibinfo{author}{Martinelli, M.},
  \bibinfo{author}{Belardinelli, A.}, \bibinfo{author}{Sparnocchia, S.},
  \bibinfo{year}{2016}.
\newblock \bibinfo{title}{Assimilation experiments for the fishery observing
  system in the adriatic sea}.
\newblock \bibinfo{journal}{Journal of Marine Systems} \bibinfo{volume}{162},
  \bibinfo{pages}{126--136}.
\bibitem[{Hernandez-Lasheras and
  Mourre(2018)}]{Hernandez-Lasheras2018DenseSardinia}
\bibinfo{author}{Hernandez-Lasheras, J.}, \bibinfo{author}{Mourre, B.},
  \bibinfo{year}{2018}.
\newblock \bibinfo{title}{{Dense CTD survey versus glider fleet sampling:
  comparison of the performance for regional ocean prediction West of
  Sardinia}}.
\newblock \bibinfo{journal}{Ocean Science Discussions} ,
  \bibinfo{pages}{1--21}\DOIprefix\doi{10.5194/os-2018-38}.
\bibitem[{Hernández-Carrasco and
  Orfila(2018)}]{https://doi.org/10.1029/2017JC013613}
\bibinfo{author}{Hernández-Carrasco, I.}, \bibinfo{author}{Orfila, A.},
  \bibinfo{year}{2018}.
\newblock \bibinfo{title}{The role of an intense front on the connectivity of
  the western mediterranean sea: The cartagena-tenes front}.
\newblock \bibinfo{journal}{Journal of Geophysical Research: Oceans}
  \bibinfo{volume}{123}, \bibinfo{pages}{4398--4422}.
\newblock \DOIprefix\doi{10.1029/2017JC013613}.
\bibitem[{Heslop et~al.(2012)Heslop, Ruiz, Allen, L{\'{o}}pez-jurado, Renault
  and Tintor{\'{e}}}]{Heslop2012}
\bibinfo{author}{Heslop, E.E.}, \bibinfo{author}{Ruiz, S.},
  \bibinfo{author}{Allen, J.}, \bibinfo{author}{L{\'{o}}pez-jurado, J.L.},
  \bibinfo{author}{Renault, L.}, \bibinfo{author}{Tintor{\'{e}}, J.},
  \bibinfo{year}{2012}.
\newblock \bibinfo{title}{{Autonomous underwater gliders monitoring variability
  at “choke points” in our ocean system: A case study in the Western
  Mediterranean Sea}}.
\newblock \bibinfo{journal}{Geophysical Research Letters} \bibinfo{volume}{39},
  \bibinfo{pages}{1--6}.
\newblock \DOIprefix\doi{10.1029/2012GL053717}.
\bibitem[{Hoffman and Atlas(2016)}]{hoffman2016future}
\bibinfo{author}{Hoffman, R.N.}, \bibinfo{author}{Atlas, R.},
  \bibinfo{year}{2016}.
\newblock \bibinfo{title}{Future observing system simulation experiments}.
\newblock \bibinfo{journal}{Bulletin of the American Meteorological Society}
  \bibinfo{volume}{97}, \bibinfo{pages}{1601--1616}.
\bibitem[{Juza et~al.(2016)Juza, Mourre, Renault, G{\'{o}}mara, Sebastian,
  L{\'{o}}pez, Borrueco, Beltran, Troupin, Tom{\'{a}}s, Heslop, Casas and
  Tintor{\'{e}}}]{Juza2016}
\bibinfo{author}{Juza, M.}, \bibinfo{author}{Mourre, B.},
  \bibinfo{author}{Renault, L.}, \bibinfo{author}{G{\'{o}}mara, S.},
  \bibinfo{author}{Sebastian, K.}, \bibinfo{author}{L{\'{o}}pez, S.L.},
  \bibinfo{author}{Borrueco, B.F.}, \bibinfo{author}{Beltran, J.P.},
  \bibinfo{author}{Troupin, C.}, \bibinfo{author}{Tom{\'{a}}s, M.T.},
  \bibinfo{author}{Heslop, E.}, \bibinfo{author}{Casas, B.},
  \bibinfo{author}{Tintor{\'{e}}, J.}, \bibinfo{year}{2016}.
\newblock \bibinfo{title}{{Operational SOCIB forecasting system and
  multi-platform validation in the Western Mediterranean}}.
\newblock \bibinfo{journal}{Journal of Operational Oceanography} ,
  \bibinfo{pages}{9231}\DOIprefix\doi{10.1002/2013JC009231.}
\bibitem[{Kai et~al.(2009)Kai, Rossi, Sudre, Weimerskirch, Lopez,
  Hernandez-Garcia, Marsac and Gar{\c{c}}on}]{kai2009top}
\bibinfo{author}{Kai, E.T.}, \bibinfo{author}{Rossi, V.},
  \bibinfo{author}{Sudre, J.}, \bibinfo{author}{Weimerskirch, H.},
  \bibinfo{author}{Lopez, C.}, \bibinfo{author}{Hernandez-Garcia, E.},
  \bibinfo{author}{Marsac, F.}, \bibinfo{author}{Gar{\c{c}}on, V.},
  \bibinfo{year}{2009}.
\newblock \bibinfo{title}{Top marine predators track lagrangian coherent
  structures}.
\newblock \bibinfo{journal}{Proceedings of the National Academy of Sciences}
  \bibinfo{volume}{106}, \bibinfo{pages}{8245--8250}.
\bibitem[{Kourafalou et~al.(2015)Kourafalou, De~Mey, Le~H{\'{e}}naff, Charria,
  Edwards, He, Herzfeld, Pascual, Stanev, Tintor{\'{e}}, Usui, van~der
  Westhuysen, Wilkin and Zhu}]{Kourafalou2015}
\bibinfo{author}{Kourafalou, V.}, \bibinfo{author}{De~Mey, P.},
  \bibinfo{author}{Le~H{\'{e}}naff, M.}, \bibinfo{author}{Charria, G.},
  \bibinfo{author}{Edwards, C.}, \bibinfo{author}{He, R.},
  \bibinfo{author}{Herzfeld, M.}, \bibinfo{author}{Pascual, a.},
  \bibinfo{author}{Stanev, E.}, \bibinfo{author}{Tintor{\'{e}}, J.},
  \bibinfo{author}{Usui, N.}, \bibinfo{author}{van~der Westhuysen, a.},
  \bibinfo{author}{Wilkin, J.}, \bibinfo{author}{Zhu, X.},
  \bibinfo{year}{2015}.
\newblock \bibinfo{title}{{Coastal Ocean Forecasting: system integration and
  evaluation}}.
\newblock \bibinfo{journal}{Journal of Operational Oceanography}
  \bibinfo{volume}{8}, \bibinfo{pages}{s127--s146}.
\newblock \DOIprefix\doi{10.1080/1755876X.2015.1022336}.
\bibitem[{Lange and Van~Sebille(2017)}]{Lange2017ParcelsAge}
\bibinfo{author}{Lange, M.}, \bibinfo{author}{Van~Sebille, E.},
  \bibinfo{year}{2017}.
\newblock \bibinfo{title}{{Parcels v0.9: Prototyping a Lagrangian Ocean
  Analysis framework for the petascale age}}.
\newblock \bibinfo{journal}{arXiv} ,
  \bibinfo{pages}{4175--4186}\DOIprefix\doi{10.5194/gmd-2017-167}.
\bibitem[{Le~Traon(2013)}]{LeTraon2013}
\bibinfo{author}{Le~Traon, P.Y.}, \bibinfo{year}{2013}.
\newblock \bibinfo{title}{{From satellite altimetry to Argo and operational
  oceanography: Three revolutions in oceanography}}.
\newblock \bibinfo{journal}{Ocean Science} \bibinfo{volume}{9},
  \bibinfo{pages}{901--915}.
\newblock \DOIprefix\doi{10.5194/os-9-901-2013}.
\bibitem[{Le~Traon et~al.(2019)Le~Traon, Reppucci, Alvarez~Fanjul, Aouf,
  Behrens, Belmonte, Bentamy, Bertino, Brando, Kreiner, Benkiran, Carval,
  Ciliberti, Claustre, Clementi, Coppini, Cossarini, De~Alfonso
  Alonso-Mu{\~{n}}oyerro, Delamarche, Dibarboure, Dinessen, Drevillon, Drillet,
  Faugere, Fern{\'{a}}ndez, Fleming, Garcia-Hermosa, Sotillo, Garric, Gasparin,
  Giordan, Gehlen, Gregoire, Guinehut, Hamon, Harris, Hernandez, Hinkler,
  Hoyer, Karvonen, Kay, King, Lavergne, Lemieux-Dudon, Lima, Mao, Martin,
  Masina, Melet, Buongiorno~Nardelli, Nolan, Pascual, Pistoia, Palazov, Piolle,
  Pujol, Pequignet, Peneva, P{\'{e}}rez~G{\'{o}}mez, Petit de~la Villeon,
  Pinardi, Pisano, Pouliquen, Reid, Remy, Santoleri, Siddorn, She, Staneva,
  Stoffelen, Tonani, Vandenbulcke, von Schuckmann, Volpe, Wettre and
  Zacharioudaki}]{LeTraon2019FromPerspective}
\bibinfo{author}{Le~Traon, P.Y.}, \bibinfo{author}{Reppucci, A.},
  \bibinfo{author}{Alvarez~Fanjul, E.}, \bibinfo{author}{Aouf, L.},
  \bibinfo{author}{Behrens, A.}, \bibinfo{author}{Belmonte, M.},
  \bibinfo{author}{Bentamy, A.}, \bibinfo{author}{Bertino, L.},
  \bibinfo{author}{Brando, V.E.}, \bibinfo{author}{Kreiner, M.B.},
  \bibinfo{author}{Benkiran, M.}, \bibinfo{author}{Carval, T.},
  \bibinfo{author}{Ciliberti, S.A.}, \bibinfo{author}{Claustre, H.},
  \bibinfo{author}{Clementi, E.}, \bibinfo{author}{Coppini, G.},
  \bibinfo{author}{Cossarini, G.}, \bibinfo{author}{De~Alfonso
  Alonso-Mu{\~{n}}oyerro, M.}, \bibinfo{author}{Delamarche, A.},
  \bibinfo{author}{Dibarboure, G.}, \bibinfo{author}{Dinessen, F.},
  \bibinfo{author}{Drevillon, M.}, \bibinfo{author}{Drillet, Y.},
  \bibinfo{author}{Faugere, Y.}, \bibinfo{author}{Fern{\'{a}}ndez, V.},
  \bibinfo{author}{Fleming, A.}, \bibinfo{author}{Garcia-Hermosa, M.I.},
  \bibinfo{author}{Sotillo, M.G.}, \bibinfo{author}{Garric, G.},
  \bibinfo{author}{Gasparin, F.}, \bibinfo{author}{Giordan, C.},
  \bibinfo{author}{Gehlen, M.}, \bibinfo{author}{Gregoire, M.L.},
  \bibinfo{author}{Guinehut, S.}, \bibinfo{author}{Hamon, M.},
  \bibinfo{author}{Harris, C.}, \bibinfo{author}{Hernandez, F.},
  \bibinfo{author}{Hinkler, J.B.}, \bibinfo{author}{Hoyer, J.},
  \bibinfo{author}{Karvonen, J.}, \bibinfo{author}{Kay, S.},
  \bibinfo{author}{King, R.}, \bibinfo{author}{Lavergne, T.},
  \bibinfo{author}{Lemieux-Dudon, B.}, \bibinfo{author}{Lima, L.},
  \bibinfo{author}{Mao, C.}, \bibinfo{author}{Martin, M.J.},
  \bibinfo{author}{Masina, S.}, \bibinfo{author}{Melet, A.},
  \bibinfo{author}{Buongiorno~Nardelli, B.}, \bibinfo{author}{Nolan, G.},
  \bibinfo{author}{Pascual, A.}, \bibinfo{author}{Pistoia, J.},
  \bibinfo{author}{Palazov, A.}, \bibinfo{author}{Piolle, J.F.},
  \bibinfo{author}{Pujol, M.I.}, \bibinfo{author}{Pequignet, A.C.},
  \bibinfo{author}{Peneva, E.}, \bibinfo{author}{P{\'{e}}rez~G{\'{o}}mez, B.},
  \bibinfo{author}{Petit de~la Villeon, L.}, \bibinfo{author}{Pinardi, N.},
  \bibinfo{author}{Pisano, A.}, \bibinfo{author}{Pouliquen, S.},
  \bibinfo{author}{Reid, R.}, \bibinfo{author}{Remy, E.},
  \bibinfo{author}{Santoleri, R.}, \bibinfo{author}{Siddorn, J.},
  \bibinfo{author}{She, J.}, \bibinfo{author}{Staneva, J.},
  \bibinfo{author}{Stoffelen, A.}, \bibinfo{author}{Tonani, M.},
  \bibinfo{author}{Vandenbulcke, L.}, \bibinfo{author}{von Schuckmann, K.},
  \bibinfo{author}{Volpe, G.}, \bibinfo{author}{Wettre, C.},
  \bibinfo{author}{Zacharioudaki, A.}, \bibinfo{year}{2019}.
\newblock \bibinfo{title}{{From Observation to Information and Users: The
  Copernicus Marine Service Perspective}}.
\newblock \bibinfo{journal}{Frontiers in Marine Science} \bibinfo{volume}{6}.
\newblock \DOIprefix\doi{10.3389/fmars.2019.00234}.
\bibitem[{Lehahn et~al.(2007)Lehahn, d'Ovidio, L{\'{e}}vy and
  Heifetz}]{Lehahn2007StirringData}
\bibinfo{author}{Lehahn, Y.}, \bibinfo{author}{d'Ovidio, F.},
  \bibinfo{author}{L{\'{e}}vy, M.}, \bibinfo{author}{Heifetz, E.},
  \bibinfo{year}{2007}.
\newblock \bibinfo{title}{{Stirring of the northeast Atlantic spring bloom: A
  Lagrangian analysis based on multisatellite data}}.
\newblock \bibinfo{journal}{Journal of Geophysical Research: Oceans}
  \bibinfo{volume}{112}, \bibinfo{pages}{1--15}.
\newblock \DOIprefix\doi{10.1029/2006JC003927}.
\bibitem[{Lekien et~al.(2005)Lekien, Coulliette, Mariano, Ryan, Shay, Haller
  and Marsden}]{lekien2005pollution}
\bibinfo{author}{Lekien, F.}, \bibinfo{author}{Coulliette, C.},
  \bibinfo{author}{Mariano, A.J.}, \bibinfo{author}{Ryan, E.H.},
  \bibinfo{author}{Shay, L.K.}, \bibinfo{author}{Haller, G.},
  \bibinfo{author}{Marsden, J.}, \bibinfo{year}{2005}.
\newblock \bibinfo{title}{Pollution release tied to invariant manifolds: A case
  study for the coast of florida}.
\newblock \bibinfo{journal}{Physica D: Nonlinear Phenomena}
  \bibinfo{volume}{210}, \bibinfo{pages}{1--20}.
\bibitem[{Moltmann et~al.(2019)Moltmann, Turton, Zhang, Nolan, Gouldman,
  Griesbauer, Willis, Piniella, Barrell, Andersson, Gallage, Charpentier,
  Belbeoch, Poli, Rea, Burger, Legler, Lumpkin, Meinig, O'Brien, Saha, Sutton,
  Zhang and Zhang}]{Moltmann2019ATechnologies}
\bibinfo{author}{Moltmann, T.}, \bibinfo{author}{Turton, J.},
  \bibinfo{author}{Zhang, H.M.}, \bibinfo{author}{Nolan, G.},
  \bibinfo{author}{Gouldman, C.}, \bibinfo{author}{Griesbauer, L.},
  \bibinfo{author}{Willis, Z.}, \bibinfo{author}{Piniella, A.M.},
  \bibinfo{author}{Barrell, S.}, \bibinfo{author}{Andersson, E.},
  \bibinfo{author}{Gallage, C.}, \bibinfo{author}{Charpentier, E.},
  \bibinfo{author}{Belbeoch, M.}, \bibinfo{author}{Poli, P.},
  \bibinfo{author}{Rea, A.}, \bibinfo{author}{Burger, E.F.},
  \bibinfo{author}{Legler, D.M.}, \bibinfo{author}{Lumpkin, R.},
  \bibinfo{author}{Meinig, C.}, \bibinfo{author}{O'Brien, K.},
  \bibinfo{author}{Saha, K.}, \bibinfo{author}{Sutton, A.},
  \bibinfo{author}{Zhang, D.}, \bibinfo{author}{Zhang, Y.},
  \bibinfo{year}{2019}.
\newblock \bibinfo{title}{{A Global Ocean Observing System (GOOS), delivered
  through enhanced collaboration across regions, communities, and new
  technologies}}.
\newblock \bibinfo{journal}{Frontiers in Marine Science} \bibinfo{volume}{6},
  \bibinfo{pages}{1--21}.
\newblock \DOIprefix\doi{10.3389/fmars.2019.00291}.
\bibitem[{Mourre et~al.(2018)Mourre, Aguiar, Juza, Hernandez-Lasheras, Reyes,
  Heslop, Escudier, Cutolo, Ruiz, Mason, Pascual and
  Tintor{\'{e}}}]{Mourre2018AssessmentSea}
\bibinfo{author}{Mourre, B.}, \bibinfo{author}{Aguiar, E.},
  \bibinfo{author}{Juza, M.}, \bibinfo{author}{Hernandez-Lasheras, J.},
  \bibinfo{author}{Reyes, E.}, \bibinfo{author}{Heslop, E.},
  \bibinfo{author}{Escudier, R.}, \bibinfo{author}{Cutolo, E.},
  \bibinfo{author}{Ruiz, S.}, \bibinfo{author}{Mason, E.},
  \bibinfo{author}{Pascual, A.}, \bibinfo{author}{Tintor{\'{e}}, J.},
  \bibinfo{year}{2018}.
\newblock \bibinfo{title}{{Assessment of High-Resolution Regional Ocean
  Prediction Systems Using Multi-Platform Observations: Illustrations in the
  Western Mediterranean Sea}}.
\newblock \bibinfo{journal}{New Frontiers in Operational Oceanography} ,
  \bibinfo{pages}{663--694}\DOIprefix\doi{10.17125/gov2018.ch24}.
\bibitem[{Pascual et~al.(2013)Pascual, Bouffard, Ruiz, {Buongiorno Nardelli},
  Vidal-Vijande, Escudier, Sayol and Orfila}]{Pascual2013}
\bibinfo{author}{Pascual, A.}, \bibinfo{author}{Bouffard, J.},
  \bibinfo{author}{Ruiz, S.}, \bibinfo{author}{{Buongiorno Nardelli}, B.},
  \bibinfo{author}{Vidal-Vijande, E.}, \bibinfo{author}{Escudier, R.},
  \bibinfo{author}{Sayol, J.M.}, \bibinfo{author}{Orfila, A.},
  \bibinfo{year}{2013}.
\newblock \bibinfo{title}{{Recent improvements in mesoscale characterization of
  the western Mediterranean Sea: synergy between satellite altimetry and other
  observational approaches}}.
\newblock \bibinfo{journal}{Scientia Marina} \bibinfo{volume}{77},
  \bibinfo{pages}{19--36}.
\newblock \DOIprefix\doi{10.3989/scimar.03740.15A}.
\bibitem[{Pinot et~al.(1994)Pinot, Tintor{\'{e}} and Gomis}]{Pinot1994}
\bibinfo{author}{Pinot, J.M.}, \bibinfo{author}{Tintor{\'{e}}, J.},
  \bibinfo{author}{Gomis, D.}, \bibinfo{year}{1994}.
\newblock \bibinfo{title}{{Quasi-synoptic mesoscale variability in the Balearic
  Sea}}.
\newblock \bibinfo{journal}{Deep-Sea Research Part I} \bibinfo{volume}{41},
  \bibinfo{pages}{897--914}.
\newblock \DOIprefix\doi{10.1016/0967-0637(94)90082-5}.
\bibitem[{Pinot et~al.(1995)Pinot, Tintor{\'{e}} and Gomis}]{Pinot1995}
\bibinfo{author}{Pinot, J.M.}, \bibinfo{author}{Tintor{\'{e}}, J.},
  \bibinfo{author}{Gomis, D.}, \bibinfo{year}{1995}.
\newblock \bibinfo{title}{{Multivariate analysis of the surface circulation in
  the Balearic Sea}}.
\newblock \bibinfo{journal}{Progress in Oceanography} \bibinfo{volume}{36},
  \bibinfo{pages}{343--376}.
\newblock \DOIprefix\doi{10.1016/0079-6611(96)00003-1}.
\bibitem[{Ryabinin et~al.(2019)Ryabinin, Barbi{\`{e}}re, Haugan, Kullenberg,
  Smith, McLean, Troisi, Fischer, Aric{\`{o}}, Aarup, Pissierssens, Visbeck,
  Enevoldsen and Rigaud}]{Ryabinin2019TheDevelopment}
\bibinfo{author}{Ryabinin, V.}, \bibinfo{author}{Barbi{\`{e}}re, J.},
  \bibinfo{author}{Haugan, P.}, \bibinfo{author}{Kullenberg, G.},
  \bibinfo{author}{Smith, N.}, \bibinfo{author}{McLean, C.},
  \bibinfo{author}{Troisi, A.}, \bibinfo{author}{Fischer, A.S.},
  \bibinfo{author}{Aric{\`{o}}, S.}, \bibinfo{author}{Aarup, T.},
  \bibinfo{author}{Pissierssens, P.}, \bibinfo{author}{Visbeck, M.},
  \bibinfo{author}{Enevoldsen, H.}, \bibinfo{author}{Rigaud, J.},
  \bibinfo{year}{2019}.
\newblock \bibinfo{title}{{The UN decade of ocean science for sustainable
  development}}.
\newblock \bibinfo{journal}{Frontiers in Marine Science} \bibinfo{volume}{6}.
\newblock \DOIprefix\doi{10.3389/fmars.2019.00470}.
\bibitem[{Sakov and Oke(2008)}]{sakov2008osse}
\bibinfo{author}{Sakov, P.}, \bibinfo{author}{Oke, P.R.}, \bibinfo{year}{2008}.
\newblock \bibinfo{title}{Objective array design: Application to the tropical
  indian ocean}.
\newblock \bibinfo{journal}{Journal of atmospheric and oceanic technology}
  \bibinfo{volume}{25}, \bibinfo{pages}{794--807}.
\bibitem[{Shadden et~al.(2005)Shadden, Lekien and
  Marsden}]{shadden2005definition}
\bibinfo{author}{Shadden, S.C.}, \bibinfo{author}{Lekien, F.},
  \bibinfo{author}{Marsden, J.E.}, \bibinfo{year}{2005}.
\newblock \bibinfo{title}{Definition and properties of lagrangian coherent
  structures from finite-time lyapunov exponents in two-dimensional aperiodic
  flows}.
\newblock \bibinfo{journal}{Physica D: Nonlinear Phenomena}
  \bibinfo{volume}{212}, \bibinfo{pages}{271--304}.
\bibitem[{Shchepetkin and McWilliams(2005)}]{Shchepetkin2005}
\bibinfo{author}{Shchepetkin, A.F.}, \bibinfo{author}{McWilliams, J.C.},
  \bibinfo{year}{2005}.
\newblock \bibinfo{title}{{The regional oceanic modeling system (ROMS): A
  split-explicit, free-surface, topography-following-coordinate oceanic
  model}}.
\newblock \bibinfo{journal}{Ocean Modelling} \bibinfo{volume}{9},
  \bibinfo{pages}{347--404}.
\newblock \DOIprefix\doi{10.1016/j.ocemod.2004.08.002}.
\bibitem[{Simoncelli et~al.(2017)Simoncelli, Fratianni, Pinardi, Grandi, Drudi,
  Oddo and Dobricic}]{CMEMS}
\bibinfo{author}{Simoncelli, S.}, \bibinfo{author}{Fratianni, C.},
  \bibinfo{author}{Pinardi, N.}, \bibinfo{author}{Grandi, A.},
  \bibinfo{author}{Drudi, M.}, \bibinfo{author}{Oddo, P.},
  \bibinfo{author}{Dobricic, S.}, \bibinfo{year}{2017}.
\newblock \bibinfo{title}{{Mediterranean Sea physical reanalysis (MEDREA
  1987-2015) (Version 1)[Data set]. Copernicus Monitoring Environment Marine
  Service (CMEMS)}}
  \DOIprefix\doi{10.25423/medsea{\_}reanalysis{\_}phys{\_}006{\_}004}.
\bibitem[{Sloyan et~al.(2019)Sloyan, Wilkin, Hill, Chidichimo, Cronin,
  Johannessen, Karstensen, Krug, Lee, Oka, Palmer, Rabe, Speich,
  Von~Schuckmann, Weller and Yu}]{Sloyan2019EvolvingCoordinatio}
\bibinfo{author}{Sloyan, B.M.}, \bibinfo{author}{Wilkin, J.},
  \bibinfo{author}{Hill, K.L.}, \bibinfo{author}{Chidichimo, M.P.},
  \bibinfo{author}{Cronin, M.F.}, \bibinfo{author}{Johannessen, J.A.},
  \bibinfo{author}{Karstensen, J.}, \bibinfo{author}{Krug, M.},
  \bibinfo{author}{Lee, T.}, \bibinfo{author}{Oka, E.},
  \bibinfo{author}{Palmer, M.D.}, \bibinfo{author}{Rabe, B.},
  \bibinfo{author}{Speich, S.}, \bibinfo{author}{Von~Schuckmann, K.},
  \bibinfo{author}{Weller, R.A.}, \bibinfo{author}{Yu, W.},
  \bibinfo{year}{2019}.
\newblock \bibinfo{title}{{Evolving the global ocean observing system for
  research and application services through international coordinatio}}.
\newblock \bibinfo{journal}{Frontiers in Marine Science} \bibinfo{volume}{6}.
\newblock \DOIprefix\doi{10.3389/fmars.2019.00449}.
\bibitem[{Smith and Sandwell(1997)}]{Smith1997GlobalSoundings}
\bibinfo{author}{Smith, W.H.}, \bibinfo{author}{Sandwell, D.T.},
  \bibinfo{year}{1997}.
\newblock \bibinfo{title}{{Global Sea Floor Topography from SAtellite Altimetry
  and Ship Depth Soundings}}.
\newblock \bibinfo{journal}{Science} \bibinfo{volume}{2}, \bibinfo{pages}{209
  --215}.
\bibitem[{Tintor{\'{e}} et~al.(2012)Tintor{\'{e}}, Lana, Marmain,
  Fern{\'{a}}ndez, Casas and Reyes}]{tintore2012hf}
\bibinfo{author}{Tintor{\'{e}}, J.}, \bibinfo{author}{Lana, A.},
  \bibinfo{author}{Marmain, J.}, \bibinfo{author}{Fern{\'{a}}ndez, V.},
  \bibinfo{author}{Casas, B.}, \bibinfo{author}{Reyes, E.},
  \bibinfo{year}{2012}.
\newblock \bibinfo{title}{{HF Radar Ibiza data from date 2012-06-01}}.
\newblock \DOIprefix\doi{https://doi.org/10.25704/17GS-2B59}.
\bibitem[{Tintor{\'{e}} et~al.(2019)Tintor{\'{e}}, Pinardi,
  {\'{A}}lvarez-Fanjul, Aguiar, {\'{A}}lvarez-Berastegui, Bajo, Balbin,
  Bozzano, Nardelli, Cardin, Casas, Charcos-Llorens, Chiggiato, Clementi,
  Coppini, Coppola, Cossarini, Deidun, Deudero, D'Ortenzio, Drago, Drudi,
  El~Serafy, Escudier, Farcy, Federico, Fern{\'{a}}ndez, Ferrarin, Fossi,
  Frangoulis, Galgani, Gana, Garc{\'{i}}a~Lafuente, Sotillo, Garreau, Gertman,
  G{\'{o}}mez-Pujol, Grandi, Hayes, Hern{\'{a}}ndez-Lasheras, Herut, Heslop,
  Hilmi, Juza, Kallos, Korres, Lecci, Lazzari, Lorente, Liubartseva, Louanchi,
  Malacic, Mannarini, March, Marullo, Mauri, Meszaros, Mourre, Mortier,
  Mu{\~{n}}oz-Mas, Novellino, Obaton, Orfila, Pascual, Pensieri,
  P{\'{e}}rez~G{\'{o}}mez, P{\'{e}}rez~Rubio, Perivoliotis, Petihakis, de~la
  Vill{\'{e}}on, Pistoia, Poulain, Pouliquen, Prieto, Raimbault, Reglero,
  Reyes, Rotllan, Ruiz, Ruiz, Ruiz, Ruiz-Orej{\'{o}}n, Salihoglu, Salon,
  Sammartino, S{\'{a}}nchez~Arcilla, Sannino, Sannino, Santoleri, Sard{\'{a}},
  Schroeder, Simoncelli, Sofianos, Sylaios, Tanhua, Teruzzi, Testor, Tezcan,
  Torner, Trotta, Umgiesser, von Schuckmann, Verri, Vilibic, Yucel, Zavatarelli
  and Zodiatis}]{Tintore2019ChallengesSea}
\bibinfo{author}{Tintor{\'{e}}, J.}, \bibinfo{author}{Pinardi, N.},
  \bibinfo{author}{{\'{A}}lvarez-Fanjul, E.}, \bibinfo{author}{Aguiar, E.},
  \bibinfo{author}{{\'{A}}lvarez-Berastegui, D.}, \bibinfo{author}{Bajo, M.},
  \bibinfo{author}{Balbin, R.}, \bibinfo{author}{Bozzano, R.},
  \bibinfo{author}{Nardelli, B.B.}, \bibinfo{author}{Cardin, V.},
  \bibinfo{author}{Casas, B.}, \bibinfo{author}{Charcos-Llorens, M.},
  \bibinfo{author}{Chiggiato, J.}, \bibinfo{author}{Clementi, E.},
  \bibinfo{author}{Coppini, G.}, \bibinfo{author}{Coppola, L.},
  \bibinfo{author}{Cossarini, G.}, \bibinfo{author}{Deidun, A.},
  \bibinfo{author}{Deudero, S.}, \bibinfo{author}{D'Ortenzio, F.},
  \bibinfo{author}{Drago, A.}, \bibinfo{author}{Drudi, M.},
  \bibinfo{author}{El~Serafy, G.}, \bibinfo{author}{Escudier, R.},
  \bibinfo{author}{Farcy, P.}, \bibinfo{author}{Federico, I.},
  \bibinfo{author}{Fern{\'{a}}ndez, J.G.}, \bibinfo{author}{Ferrarin, C.},
  \bibinfo{author}{Fossi, C.}, \bibinfo{author}{Frangoulis, C.},
  \bibinfo{author}{Galgani, F.}, \bibinfo{author}{Gana, S.},
  \bibinfo{author}{Garc{\'{i}}a~Lafuente, J.}, \bibinfo{author}{Sotillo, M.G.},
  \bibinfo{author}{Garreau, P.}, \bibinfo{author}{Gertman, I.},
  \bibinfo{author}{G{\'{o}}mez-Pujol, L.}, \bibinfo{author}{Grandi, A.},
  \bibinfo{author}{Hayes, D.}, \bibinfo{author}{Hern{\'{a}}ndez-Lasheras, J.},
  \bibinfo{author}{Herut, B.}, \bibinfo{author}{Heslop, E.},
  \bibinfo{author}{Hilmi, K.}, \bibinfo{author}{Juza, M.},
  \bibinfo{author}{Kallos, G.}, \bibinfo{author}{Korres, G.},
  \bibinfo{author}{Lecci, R.}, \bibinfo{author}{Lazzari, P.},
  \bibinfo{author}{Lorente, P.}, \bibinfo{author}{Liubartseva, S.},
  \bibinfo{author}{Louanchi, F.}, \bibinfo{author}{Malacic, V.},
  \bibinfo{author}{Mannarini, G.}, \bibinfo{author}{March, D.},
  \bibinfo{author}{Marullo, S.}, \bibinfo{author}{Mauri, E.},
  \bibinfo{author}{Meszaros, L.}, \bibinfo{author}{Mourre, B.},
  \bibinfo{author}{Mortier, L.}, \bibinfo{author}{Mu{\~{n}}oz-Mas, C.},
  \bibinfo{author}{Novellino, A.}, \bibinfo{author}{Obaton, D.},
  \bibinfo{author}{Orfila, A.}, \bibinfo{author}{Pascual, A.},
  \bibinfo{author}{Pensieri, S.}, \bibinfo{author}{P{\'{e}}rez~G{\'{o}}mez,
  B.}, \bibinfo{author}{P{\'{e}}rez~Rubio, S.}, \bibinfo{author}{Perivoliotis,
  L.}, \bibinfo{author}{Petihakis, G.}, \bibinfo{author}{de~la Vill{\'{e}}on,
  L.P.}, \bibinfo{author}{Pistoia, J.}, \bibinfo{author}{Poulain, P.M.},
  \bibinfo{author}{Pouliquen, S.}, \bibinfo{author}{Prieto, L.},
  \bibinfo{author}{Raimbault, P.}, \bibinfo{author}{Reglero, P.},
  \bibinfo{author}{Reyes, E.}, \bibinfo{author}{Rotllan, P.},
  \bibinfo{author}{Ruiz, S.}, \bibinfo{author}{Ruiz, J.},
  \bibinfo{author}{Ruiz, I.}, \bibinfo{author}{Ruiz-Orej{\'{o}}n, L.F.},
  \bibinfo{author}{Salihoglu, B.}, \bibinfo{author}{Salon, S.},
  \bibinfo{author}{Sammartino, S.}, \bibinfo{author}{S{\'{a}}nchez~Arcilla,
  A.}, \bibinfo{author}{Sannino, G.}, \bibinfo{author}{Sannino, G.},
  \bibinfo{author}{Santoleri, R.}, \bibinfo{author}{Sard{\'{a}}, R.},
  \bibinfo{author}{Schroeder, K.}, \bibinfo{author}{Simoncelli, S.},
  \bibinfo{author}{Sofianos, S.}, \bibinfo{author}{Sylaios, G.},
  \bibinfo{author}{Tanhua, T.}, \bibinfo{author}{Teruzzi, A.},
  \bibinfo{author}{Testor, P.}, \bibinfo{author}{Tezcan, D.},
  \bibinfo{author}{Torner, M.}, \bibinfo{author}{Trotta, F.},
  \bibinfo{author}{Umgiesser, G.}, \bibinfo{author}{von Schuckmann, K.},
  \bibinfo{author}{Verri, G.}, \bibinfo{author}{Vilibic, I.},
  \bibinfo{author}{Yucel, M.}, \bibinfo{author}{Zavatarelli, M.},
  \bibinfo{author}{Zodiatis, G.}, \bibinfo{year}{2019}.
\newblock \bibinfo{title}{{Challenges for Sustained Observing and Forecasting
  Systems in the Mediterranean Sea}}.
\newblock \bibinfo{journal}{Frontiers in Marine Science} \bibinfo{volume}{6}.
\newblock \DOIprefix\doi{10.3389/fmars.2019.00568}.
\bibitem[{Und{\'{e}}n et~al.(2002)Und{\'{e}}n, Rontu, J{\"{a}}rvinen, Lynch,
  Calvo, Cats, Cuxart, Eerola, Fortelius, Garcia-Moya, Jones, Lenderlink,
  McDonald, McGrath, Navascues, Nielsen, {\O}degaard, Rodriguez, Rummukainen,
  Rõõm, Sattler, Sass, Savij{\"{a}}rvi, Schreur, Sigg, The and
  Tijm}]{Unden2002HighDocumentation}
\bibinfo{author}{Und{\'{e}}n, P.}, \bibinfo{author}{Rontu, L.},
  \bibinfo{author}{J{\"{a}}rvinen, H.}, \bibinfo{author}{Lynch, P.},
  \bibinfo{author}{Calvo, J.}, \bibinfo{author}{Cats, G.},
  \bibinfo{author}{Cuxart, J.}, \bibinfo{author}{Eerola, K.},
  \bibinfo{author}{Fortelius, C.}, \bibinfo{author}{Garcia-Moya, J.A.},
  \bibinfo{author}{Jones, C.}, \bibinfo{author}{Lenderlink, G.},
  \bibinfo{author}{McDonald, A.}, \bibinfo{author}{McGrath, R.},
  \bibinfo{author}{Navascues, B.}, \bibinfo{author}{Nielsen, N.W.},
  \bibinfo{author}{{\O}degaard, V.}, \bibinfo{author}{Rodriguez, E.},
  \bibinfo{author}{Rummukainen, M.}, \bibinfo{author}{Rõõm, R.},
  \bibinfo{author}{Sattler, K.}, \bibinfo{author}{Sass, B.H.},
  \bibinfo{author}{Savij{\"{a}}rvi, H.}, \bibinfo{author}{Schreur, B.W.},
  \bibinfo{author}{Sigg, R.}, \bibinfo{author}{The, H.}, \bibinfo{author}{Tijm,
  A.}, \bibinfo{year}{2002}.
\newblock \bibinfo{title}{{High Resolution Limited Area Model, HIRLAM-5
  Scientific Documentation}}.
\newblock \bibinfo{type}{Technical Report} \bibinfo{number}{December}. SMHI.
  \bibinfo{address}{Norrkoping, SWEDEN}.
\bibitem[{Vandenbulcke et~al.(2017)Vandenbulcke, Beckers and
  Barth}]{Vandenbulcke2017}
\bibinfo{author}{Vandenbulcke, L.}, \bibinfo{author}{Beckers, J.m.},
  \bibinfo{author}{Barth, A.}, \bibinfo{year}{2017}.
\newblock \bibinfo{title}{{Correction of inertial oscillations by assimilation
  of HF radar data in a model of the Ligurian Sea}}.
\newblock \bibinfo{journal}{Ocean Dynamics} ,
  \bibinfo{pages}{117--135}\DOIprefix\doi{10.1007/s10236-016-1012-5}.
\bibitem[{Vignudelli et~al.(2019)Vignudelli, Birol, Benveniste, Fu, Picot,
  Raynal and Roinard}]{Vignudelli2019SatelliteZone}
\bibinfo{author}{Vignudelli, S.}, \bibinfo{author}{Birol, F.},
  \bibinfo{author}{Benveniste, J.}, \bibinfo{author}{Fu, L.L.},
  \bibinfo{author}{Picot, N.}, \bibinfo{author}{Raynal, M.},
  \bibinfo{author}{Roinard, H.}, \bibinfo{year}{2019}.
\newblock \bibinfo{title}{{Satellite Altimetry Measurements of Sea Level in the
  Coastal Zone}}. volume~\bibinfo{volume}{40}.
\newblock \bibinfo{publisher}{Springer Netherlands}.
\newblock \DOIprefix\doi{10.1007/s10712-019-09569-1}.
\bibitem[{Visbeck(2018)}]{Visbeck2018OceanFuture}
\bibinfo{author}{Visbeck, M.}, \bibinfo{year}{2018}.
\newblock \bibinfo{title}{{Ocean science research is key for a sustainable
  future}}.
\newblock \bibinfo{journal}{Nature Communications} \bibinfo{volume}{9},
  \bibinfo{pages}{1--4}.
\newblock \DOIprefix\doi{10.1038/s41467-018-03158-3}.

\end{thebibliography}

\end{document}